\documentclass[hidelinks,12pt]{article}
\usepackage{graphicx}
\usepackage[none]{hyphenat}
\usepackage[usenames, dvipsnames]{xcolor}
\usepackage{multirow}
\usepackage{listings}
\usepackage{setspace}
\usepackage{booktabs}
\usepackage{array}
\usepackage{multicol}
\usepackage{amssymb}
\usepackage{gensymb}
\usepackage{bm}
\usepackage{soul}
\usepackage{amsmath}
\usepackage{mathtools}
\usepackage[utf8]{inputenc}
\usepackage[english]{babel}
\usepackage{hyperref}
\usepackage{setspace}
\usepackage{parskip}
\usepackage{titlesec}
\usepackage{lineno}
\usepackage{lscape}
\usepackage{rotating}
\usepackage[headheight=15pt, letterpaper, margin=1.0in]{geometry}

\newcommand{\bt}[1]{{\color{black} #1}}
\newcommand{\rt}[1]{{\color{black} #1}}

\begin{document}
	
	
	\title{A Hierarchical Bayes Unit-Level Small Area Estimation Model for Normal Mixture Populations}
	
	\author{Shuchi Goyal\\Department of Statistics\\ University of California at Los Angeles\\ Los Angeles, CA 90095 \and Gauri Sankar Datta\\ Department of Statistics\\University of Georgia\\Athens, GA 30602  \and Abhyuday Mandal\\Department of Statistics\\University of Georgia\\Athens, GA 30602}
	\date{\today}

	
	\maketitle
	
	\begin{abstract}
		National statistical agencies are regularly required to produce estimates about various subpopulations, formed by demographic and/or geographic classifications, based on a limited number of samples. Traditional direct estimates computed using only sampled data from individual subpopulations are usually unreliable due to small sample sizes. Subpopulations with small samples are termed small areas or small domains. To improve on the less reliable direct estimates, model-based estimates, which borrow information from suitable auxiliary variables, have been extensively proposed in the literature. However, standard model-based estimates rely on the normality assumptions of the error terms. In this research we propose a hierarchical Bayesian (HB) method for the unit-level nested error regression model based on a normal mixture for the unit-level error distribution. Our method proposed here is applicable to model cases with unit-level error outliers as well as cases where each small area population is comprised of two subgroups, neither of which can be treated as an outlier. Our proposed method is more robust than the normality based standard HB method (Datta and Ghosh 1991) to handle outliers or multiple subgroups in the population. Our proposal assumes two subgroups and the two-component mixture model that has been recently proposed by Chakraborty et al. (2018) to address outliers. To implement our proposal we use a uniform prior for the regression parameters, random effects variance parameter, and the mixing proportion, and we use a partially proper non-informative prior distribution for the two unit-level error variance components in the mixture. We apply our method to two examples to predict summary characteristics of farm products at the small area level. One of the examples is prediction of twelve county-level crop areas cultivated for corn in some Iowa counties. The other example involves total cash associated in farm operations in twenty-seven farming regions in Australia. 
		
		We compare predictions of small area characteristics based on the proposed method with those obtained by applying the Datta and Ghosh (1991) and the Chakraborty et al. (2018) HB methods. 
		Our simulation study comparing these three Bayesian methods, when the unit-level error distribution is normal, or $t$, or two-component normal mixture, showed the superiority of our proposed method, measured by prediction mean squared error, coverage probabilities and lengths of credible intervals for the small area means.
	\end{abstract}
	

	
	\section{Introduction}
	\noindent National statistical offices around the world have been mandated for many years to produce reliable statistics for important variables such as population, income, unemployment, and health outcomes for various geographic domains (e.g., states, counties) and/or demographic domains (e.g., age, race, gender). However, the sample available from many of these domains are often small to produce direct estimates of adequate accuracy. This situation is known as small area estimation. To develop estimates that are more reliable than the direct estimates, data from the entire sample (that is, a sample covering all small areas) is used and combined with other appropriate auxiliary variables to produce indirect estimates of the small domain characteristics. Model-based approaches have been shown to be useful in producing reliable small area or small domain estimates. 
	
	\bigskip\noindent \bt{The earliest important application of model based small area estimation is by Fay and Herriot (1979). They adopted shrinkage estimation of Stein (\rt{Ref}), popularized as empirical Bayes estimation (Efron and Morris, 1973). Using empirical Bayes method, Fay and Herriot (1979) proposed shrinkage of a direct estimator of a small-area mean to a regression plane determined jointly by the direct estimators and suitable auxiliary variables from the small areas. This approach is based on modeling of small area level summary statistics, often sample means.
		
	\bigskip\noindent Battese, Harter, and Fuller (1988) proposed the popular nested-error regression (NER) model to develop small area estimates based on data available on the individual sampled units. Battese et al. (1988) proposed for unit-level response a regression model based on unit-level auxiliary variables. The NER model, aptly called unit-level model, is developed under the normality assumption of small area random effects and unit-level random errors. For unit-level data, the NER model has been the basis for producing reliable small-area estimates either by a frequentist or a Bayesian approach. Datta and Ghosh (1991) used the NER model, in conjunction with suitable noninformative priors for the regression coefficients and variance parameters, to develop hierarchical Bayes estimates of finite population small area means. Prasad and Rao (1990) and Datta and Lahiri (2000) used a frequentist approach for the NER model to develop empirical best linear unbiased prediction (EBLUP) of the finite population means. To facilitate our discussion of robust HB method of small area estimation we reviewed the existing HB models in the next section.}
		
	\bigskip\noindent It is desirable to have a model that is robust in the presence of random errors prone to outliers. To address the specific case where outliers are present in the data, Chakraborty, Datta, and Mandal (2018) proposed an HB alternative to Datta and Ghosh's method (1991). By using a two-component mixture of normal distribution, this model accommodates populations where a small portion of unit-level errors come from a secondary distribution with a larger variance than the primary distribution. Chakraborty et al. (2018) showed that their model consistently performs as well as or better than that of Datta and Ghosh (1991), including in the special case of no outliers (i.e. ``no contamination'').
	
	\bigskip\noindent It should be noted that the model proposed by Chakraborty et al. (2018) is most effective when only a small portion of the population comes from the secondary distribution. In this paper we suggest an HB method built from the NER model to handle more general cases of two-component mixture populations, where the proportion of members from the secondary distribution may be as high as 50 percent.

	\clearpage
	\section{Existing Unit-Level HB Small Area Models}\label{sec:currMods}

	\bt{The NER model of Battese et al. (1988) is immensely popular in unit-level modeling for small area estimation. This model supposes that a population is partitioned in $m$ small areas with $N_i$ units in the $i$th small area. The value of the response variable for the $j$th unit in the $i$th small area $Y_{ij}$ satisfies
	\begin{eqnarray}
	Y_{ij}=\bm{x}^T_{ij}\bm{\beta}+v_i+e_{ij}, j=1,\ldots,N_i, i=1,\ldots,m,
	\label{eq:BHF}
	\end{eqnarray}
	where $Y_{ij}$ is the response variable for the $j$th unit of the $i$th small area and $\bm{x_{ij}}=(x_{ij1},\ldots,x_{ijq})^T$ is a $q\times1$ vector of values for predictor variables for that observation. Here $\bm{\beta}=(\beta_1,\ldots,\beta_q)^T$ denotes the vector of regression coefficients. The {zero mean random} variables $v_i$ and $e_{ij}$ account for area- and unit-level errors, respectively, and are assumed to be independent of each other. We further assume that $v_i$'s are i.i.d. $N(0,\sigma_v^2)$. As in Battese et al. (1988), under appropriate distributional assumptions for the $e_{ij}$'s, our goal is to predict the population mean  $\theta_{i}$ in the $i$th county defined as the conditional mean of the response given the realized random effect $v_i$, where $\theta_{i} =\bar{\bm{x}}^T_{i(p)}\bm{\beta}+v_i$, and $\bar{\bm{x}}_{i(p)} = \frac{1}{N_i} \sum_{j=1}^{N_i}x_{ij}$. The $\bar{\bm{x}}_{i(p)}$'s are known for all the small areas.

	\bigskip\noindent A special case of an HB model introduced by Datta and Ghosh (1991) includes the following HB version of the NER model. We denote this by DG HB model.
	\renewcommand{\theenumi}{(\Roman{enumi})}
	\begin{enumerate}
		\item Conditional on $\bm{\beta}, \bm{v}=(v_1,...,v_m)^T,\sigma_e^2,$ and $\sigma_v^2$,
		\begin{eqnarray*}
			Y_{ij} \stackrel{ind}{\sim} N(\bm{x}_{ij}^T\bm{\beta} + v_i,\sigma_e^2)
		\end{eqnarray*}
		for $j=1,...,N_i, i=1,...,m$.
		\item Conditional on $\bm{\beta}, \sigma_e^2$ and $\sigma_v^2$, $v_i \stackrel{\text{iid}}{\sim} N(0,\sigma_v^2)$ for all $i$.
		\item Model parameters $\bm{\beta}, \sigma_e^2$ and $\sigma_v^2$ are assigned the improper prior
			\begin{eqnarray}
		\pi(\bm{\beta},\sigma_v^2, \sigma_e^2) \propto \dfrac{1}{\sigma_e^2}.
		\label{eq:DGprior}
		\end{eqnarray}
	\end{enumerate}
	Based on a random sample of size $n_i, i=1,...,m$, from all the small areas Datta and Ghosh (1991) used the above model to develop HB predictors of small area finite population means $\bar{Y}_i$'s, $i=1,...,m$. This model can also be used to develop Bayes predictors of $\theta_i$'s. For small $n_i/N_i$, the two quantities  $\bar{Y}_i$ and $\theta_i$'s are approximately the same.

	\bigskip\noindent While the HB estimates developed by Datta and Ghosh (1991) are effective for populations in which the unit-level random errors follow a normal distribution, they are less effective when the errors follow a mixture of normal distributions. This scenario can be formulated by a two-component mixture model for the unit-level errors. The model accommodates two normal distributions underlying the unit-level error term which have the same mean but different variances. Another example of this situation is a population with ``representative outliers'' (Chambers 1986). In this case, the underlying distribution of outliers is assumed to have the same mean zero, but a larger variance than that of the non-outliers.
	}
		
	\bigskip\noindent Chakraborty et al. (2018) proposed a two-component normal mixture for the unit-level error distribution. The latter model, referred to as the CDM model hereafter, specifically facilitates small area estimation for populations which are suspected to contain representative outliers. Chambers (1986) defines a representative outlier as a value which is non-unique in the population and influences the estimates of finite population means $\bar{Y}_i$'s from the model. The CDM HB model, which modifies the DG HB model, is given below: 
	\renewcommand{\theenumi}{(\Roman{enumi})}
	\begin{enumerate}
	\item Conditional on $\bm{\beta}=(\beta_1,...,\beta_q)^T,v_i,z_{ij},p_e,\sigma_1^2,\sigma_2^2,$ and $\sigma_v^2$,
	\begin{eqnarray*}
		Y_{ij} \sim z_{ij}N(\bm{x}_{ij}^T\bm{\beta} + v_i,\sigma_1^2)+(1-z_{ij})N(\bm{x}_{ij}^T\bm{\beta}+v_i,\sigma_2^2)
	\end{eqnarray*}
	for $j=1,...,N_i, i=1,...,m$.
	\item The indicator variables $z_{ij}$ are i.i.d. with $P(z_{ij}=1|p_e)=p_e$ and $P(z_{ij}=0|p_e)=1-p_e$ for all $i,j$. Also, $z_{ij}$'s are independent of $v_i$'s,  $\bm{\beta}$, $\sigma_1^2$, $\sigma_2^2$, and $\sigma_v^2$.
	\item Conditional on $\bm{\beta}, z, p_e, \sigma_1^2, \sigma_2^2,$ and $\sigma_v^2$, $v_i \stackrel{\text{iid}}{\sim} N(0,\sigma_v^2)$ for all $i$.
	\end{enumerate}
	{A key component of the CDM HB model is that outlier observations come from a distribution which has the same mean $\bm{x}_{ij}^T\bm{\beta}+v_i$ (conditional on random effects) as the distribution of non-outliers but a larger variance. The variances for non-outliers and outliers are denoted as $\sigma_1^2$ and $\sigma_2^2$, respectively, with $\sigma_1^2<\sigma_2^2$. A priori outliers are assumed to occur in the various small areas with equal probability $(1-p_e)$.} The CDM HB model is completed by assigning independent noninformative priors for $\bm{\beta}, \sigma^2_1, \sigma^2_2, p_e,\text{ and }\sigma^2_v$, with $\bm{\beta}\sim Uniform(R^q), \sigma^2_v\sim Uniform(R^+),$ $ \pi(\sigma^2_1,\sigma^2_2)\propto \dfrac{1}{(\sigma_2^2)^2}I(\sigma^2_1<\sigma^2_2),$ and $p_e\sim Uniform(0,1)$.
	
	\bigskip\noindent \bt{The DG HB model is a limiting version of the CDM HB model when $p_e$ is on the boundary. 
		
	\bigskip\noindent In the frequentist approach Prasad and Rao (1990) also used the NER model to derive the EBLUPs of $\theta_i$ and  $\bar{Y}_i$ and estimators of their mean squared errors (MSE). In a subsequence article, Sinha and Rao (2009) investigated robustness of EBLUPs and the estimates of MSE in the presence of outliers. Their investigation showed that while the departure of random small area effects from the normality does not severely affect the EBLUPs and MSE estimates, a departure of normality assumption of the unit-level error terms adversely impacts the EBLUPs and their MSE estimates. Sinha and Rao (2009) proposed a robust empirical best linear unbiased prediction (REBLUP) approach to mitigate the impact of outliers in the unit-level error and/or in the area-level random effects on the EBLUPs. Chakraborty et al. (2018) carried out simulations to show that their proposed robust Bayesian method and the REBLUP method perform very similarly. Due to a lack of space we exclude the Sinha-Rao frequentist method here to focus only on Bayesian techniques. For more details on the comparative performance of these robust methods and the DG HB method, we refer the reader to Section 6 of Chakrabarty et al. (2018).
	
	\bigskip\noindent We note that Chakraborty et al. (2018) used a Bayesian version of a popular contamination model to accommodate a small fraction of outliers in the sample. There are applications where the population is actually a mixture of multiple component distributions, where each component is a significant minority. To address such applications, we propose an HB model built from the NER model to handle more general cases of two-component mixture populations, where the proportion of members from the secondary distribution may be nearly half. The proposed new model is more appropriate to deal with mixture populations, comprised of two sub-groups, differentiated by their variances. We propose our new model in Section 3.  We apply this new HB model as well as the DG HB and CDM HB models in Section 4 to two examples to predict summary characteristics of farm products at the small area level. One of the examples is prediction of twelve county-level crop areas cultivated for corn in some Iowa counties (cf. Battese et al., 1988). The other example involves total cash associated in farm operations in twenty-seven farming regions in Australia based on a dataset by Chambers et al. (2011). In the setup of the corn data, we compare in Section 5 our proposed method with two competing Bayesian methods via simulation studies. Two data analyses and the simulation studies demonstrated the superiority of the new proposed HB model. Concluding comments are provided in Section 6, and relevant proofs and in-depth details are relegated to the Appendix and Supplementary Information sections.
	}

	\section{An HB Normal Mixture Model for Unit-Level Error}\label{sec:newMod}
	
	The proposed model is a mixture extension of the nested-error regression model which accounts for unit-level error terms coming from two different normal distributions. While the models discussed in the previous section accommodate the presence of outliers, our proposed model further generalizes the mixture model for the case in which the proportion of observations coming from the subpopulation with a larger variance is large enough that these data points may no longer be considered outliers in the traditional sense.
	
	\bigskip\noindent We first consider the general form of the nested-error regression model for unit-level data, given in Equation (1). To extend the basic NER model to account for observations from a mixture of two underlying distributions, we rely on the same assumptions (I) to (III) of the CDM model. \bt{The CDM model is a contamination model frequently used in the literature to accommodate a handful of outlying observations. In some applications, however, there may be a larger proportion of observations which may be different from the rest of the data. In these cases, since this group of observations is not really outliers, a mixture model, which we propose below, will be better suited than the contamination model. However, the proposed mixture model is also flexible enough to explain a small fraction of outliers in a dataset.}
	
	\bt{In our new formulation of the two-component mixture model for the unit-level error component, we treat the unit-level variances $\sigma_1^2$ and $\sigma_2^2$ symmetrically, and consequently assign a prior
		\begin{eqnarray*}
			\pi(\sigma_1^2,\sigma_2^2)\propto\dfrac{1}{(\sigma_1^2+\sigma_2^2)^2}.
		\end{eqnarray*}
		It is a key difference in the priors assumed in our proposed model, which we refer as GDM hereafter, and those used for CDM. It is important to ensure the identifiability of all the parameters in the likelihood of the mixture model described by the hierarchy (I) to (III). We achieve this by assuming $p_e > 2^{-1}$, that avoids the label-switching problem. To complete specification of the prior distributions for the remaining parameters, we also assign the same independent noninformative uniform priors to $\bm{\beta}$, $\sigma_v^2$ and $p_e$ given by
		\begin{eqnarray*}
			\pi(\bm{\beta},\sigma_v^2,p_e)\propto I_{(p_e \ge 2^{-1})}.
		\end{eqnarray*}
		The GDM mixture formulation presented above differs from the CDM model, which attains identifiability by constraining $\sigma_1^2<\sigma_2^2$ but not $p_e$.}

	\bigskip\noindent To proceed with Bayesian inference based on an improper prior, the propriety of the posterior distribution must be justified. This propriety is demonstrated in the Appendix A.1. Details on the procedure for Gibbs sampling for fitting the model can be found in the Supplementary Material included at the end of this paper.

	
	\section{Data Analysis}\label{sec:data}
	
	\subsection{Prediction of County Means of Crop Areas}
	Battese et al. (1988) proposed the NER model to compute EBLUP prediction of mean hectares of corn grown in twelve counties of Iowa based on auxiliary variables provided by LANDSAT satellite data from the U.S. Department of Agriculture. {The two auxiliary variables considered are mean number of pixels of corn and soybeans in sample segments satellite imaging.} Of 37 measurements of hectors of corn samples, one observation from Hardin County was suspected of being an outlier. {The reported hectares of corn in this segment seems to be very low relative to the pixels of corn observed there, relative to other segments in the same county.} Battese et al. (1988) suggested removing the suspected outlier {altogether} from the data set to improve the fit of the basic nested-error regression model. Datta and Ghosh (1991) subsequently used this reduced data to develop their HB prediction.
	
	\bigskip\noindent It is well-known that discarding suspected outliers can lead to loss of valuable information about the data set. \bt{By including the outlier from Hardin County when fitting a robust model, it would make sense that the estimated mean corn hectares would be higher than in the non-robust DG model.} Chakraborty et al. (2018) demonstrated that when the full set of sampled observations is used, their HB prediction (CDM HB) of mean in Hardin County is closer to estimates produced by the REBLUPs of Sinha and Rao (2009) to the robust EBLUP approach proposed by Sinha and Rao (2009), than the prediction obtained from the DG HB model. When applied to the reduced data set ($n=36$), where the suspected outlier is discarded, the CDM HB model performs {similarly to} the DG HB model, indicating no loss in applying the CDM model to data which may not have any outliers. We apply the proposed model to calculate point estimates (posterior means) and standard errors (posterior standard deviations) of mean corn production in each county and compare our results to the predictions obtained from DG and CDM models. The results are shown in Table 1.
	
	\bigskip\noindent Our proposed model performs as well as the CDM model in the presence of a suspected outlier. The point estimates and standard errors calculated based on the proposed model, with the exception of one county, are very close to those produced by the CDM method. While there is considerable agreement in the estimates from the two robust Bayesian methods, these estimates are substantially different from those from the non-robust DG HB method.
	
	\begin{table}[ht]\caption{Various HB point estimates and standard errors of county hectares of corn (Full) }\label{tab1Full:BHFanalysis}
		\scriptsize
		\begin{center}
			\begin{tabular}{lc|rr|rr|rr}
				\hline
				County & {$n_i$}  & \multicolumn{2}{c}{DG}   &\multicolumn{2}{c}{CDM}    & \multicolumn{2}{c}{GDM}\\
				& & Mean & SD &     Mean & SD &    Mean & SD \\
				\hline
				Cerro Gordo    & 1 &    123.8 & 11.7     &    123.4 & 9.8      &    123.6 & 11.3   \\
				Hamilton       & 1 &    124.9 & 11.4     &    126.6 & 10.3     &    125.8 & 10.2   \\
				Worth          & 1 &    110.0 & 12.3     &    108.0 & 11.3     &    107.7 & 11.7   \\ 
				Humboldt       & 2 &    114.2 & 10.7     &    112.3 & 10.2     &    112.0 & 10.7   \\  
				Franklin       & 3 &    140.3 & 10.8     &    142.1 & 8.1      &    142.4 & 8.4    \\ 
				Pocahontas     & 3 &    110.0 & 9.6      &    111.4 & 7.6      &    111.6 & 7.3    \\ 
				Winnebago      & 3 &    116.0 & 9.7      &    114.3 & 7.6      &    113.7 & 7.9    \\ 
				Wright         & 3 &    123.2 & 9.5      &    122.7 & 7.9      &    122.3 & 7.7    \\  
				Webster        & 4 &    112.6 & 9.9      &    113.9 & 6.9      &    114.3 & 6.8    \\
				Hancock        & 5 &    124.4 & 8.9      &    123.5 & 6.1      &    123.6 & 6.1    \\ 
				Kossuth        & 5 &    111.3 & 8.9      &    108.2 & 6.8      &    108.1 & 6.9    \\ 
				Hardin         & 6 &    130.7 & 8.3      &    135.3 & 7.5      &    136.5 & 7.4    \\
				\hline
			\end{tabular}
		\end{center}
	\end{table}
	
	\begin{table}[ht]\caption{Various HB estimates and standard errors of county hectares of corn (Reduced) }\label{tab1R2:BHFanalysis}
		\scriptsize
		\begin{center}
			
			\begin{tabular}{lc|rr|rr|rr}
				\hline
				SA & {$n_i$}  & \multicolumn{2}{c}{DG}   &\multicolumn{2}{c}{CDM}    & \multicolumn{2}{c}{GDM}\\
				& & Mean & SD &     Mean & SD &    Mean & SD \\
				\hline
				Cerro Gordo   & 1 &    122.0 & 11.6   & 121.7 & 9.7      &    121.9 & 10.2   \\ 
				Hamilton      & 1 &    126.4 & 10.9   & 127.2 & 9.7      &    126.3 & 9.8   \\
				Worth         & 1 &    107.6 & 12.4   & 105.6 & 10.1     &    105.3 & 10.9   \\ 
				Humboldt      & 2 &    108.9 & 10.5   & 108.2 & 8.7      &    108.0 & 9.3   \\  
				Franklin      & 3 &    143.6 & 9.7    & 144.1 & 7.0      &    144.3 & 7.0   \\ 
				Pocahontas    & 3 &    112.3 & 9.7    & 112.5 & 6.5      &    112.3 & 6.7   \\ 
				Winnebago     & 3 &    113.4 & 9.1    & 112.5 & 6.8      &    111.5 & 7.4   \\ 
				Wright        & 3 &    121.9 & 8.8    & 121.9 & 6.6      &    121.8 & 6.7   \\  
				Webster       & 4 &    115.5 & 9.2    & 115.7 & 5.7      &    115.8 & 6.1   \\
				Hancock       & 5 &    124.8 & 8.4    & 124.4 & 5.4      &    124.6 & 5.5   \\ 
				Kossuth       & 5 &    107.7 & 8.5    & 106.3 & 5.7      &    106.0 & 5.6   \\ 
				Hardin        & 5 &    142.6 & 9.0    & 143.5 & 5.9      &    143.6 & 5.6   \\
				\hline
			\end{tabular}
		\end{center}
	\end{table}
	\bigskip\noindent 
	
	\bt{We present in Figure 1 a graphical display of posterior probability that an observation's unit-level error comes from subpopulation~2. The horizontal axes of the plots in Figure 1 represent the standardized versions of the reported hectares of corn in a surveyed segment $y_{ij}$, defined by $E(y_{ij}-\theta_i|\bm{y})/\sqrt{var(y_{ij}-\theta_i|\bm{y})}$ where $\bm{y}=\{y_{ij};j=1,...,n_i, i=1,...,m\}$, $E(\cdot|\bm{y})$ and $var(\cdot|\bm{y})$ represent the posterior mean and posterior variance operators. The vertical axes represent the posterior probability of an observation coming from the subpopulation 2.} The GDM model identifies the second Hardin County observation, which has the most extreme standardized residual, as a likely member of a secondary subpopulation when analyzing the full data, as shown in the left panel of Figure 1. The posterior probability that this observation may belong to the secondary population is 0.62, which is about 2.5 times the corresponding prior probability 0.25. For the other observations, most of their posterior probabilities are near 0.25, not much different from their prior values. We note here that in the right panel of Figure 1, we plotted the same posterior probabilities for the {\it reduced} data, after removing the second observation from Hardin county. Interestingly, for none of these observations, the posterior probabilities are greater than 0.25.

	\bigskip\noindent We also compare model estimates for the data set after removing the outlier. The point estimates and posterior standard deviations given by each method for the reduced data are given in Table 2. For the first 11 counties listed, the estimates produced by each model change only slightly when compared to those calculated using the full data set. As expected, the estimate for Hardin County changes the most significantly. With the outlier removed, the point estimates for Hardin County increase in all three models but the change is less for the two mixture models and is the most substantial in the DG model. (similar conclusion was reached by Chakraborty et al. (2018) for Sinha-Rao REBLUP, see Table~1 of Chakrabarty et al. (2018).) This makes sense, as the estimates from the mixture models should be less sensitive to the presence of outliers. The corn hectare estimates in {Hardin County} given by the three models are also much closer in value to each other relative to the full data set. (Again, these estimates agree very closely with the REBLUP estimate; see Table~1 of Chakrabarty et al. (2018).).
	
	\noindent For the reduced data set, a comparison of posterior standard deviations associated with the HB estimates show that the standard deviations from the mixture models are consistently lower than those given by the DG model. We also compare posterior standard deviations between the full data analysis and the reduced data analysis. Intuitively, the presence of an outlier will cause an increase in unit-level variances, and therefore may also cause an increase in {posterior variances of small area means}. While the standard deviations produced by the robust CDM and GDM HB models appear to be higher for the full data than for the reduced data, the standard deviations given by the DG model seem to change only moderately. Standard deviations for the non-robust HB DG model are the highest.
	\begin{figure}[h]
		\begin{center}
			\begin{tabular}{cc}
				\includegraphics[scale=.35]{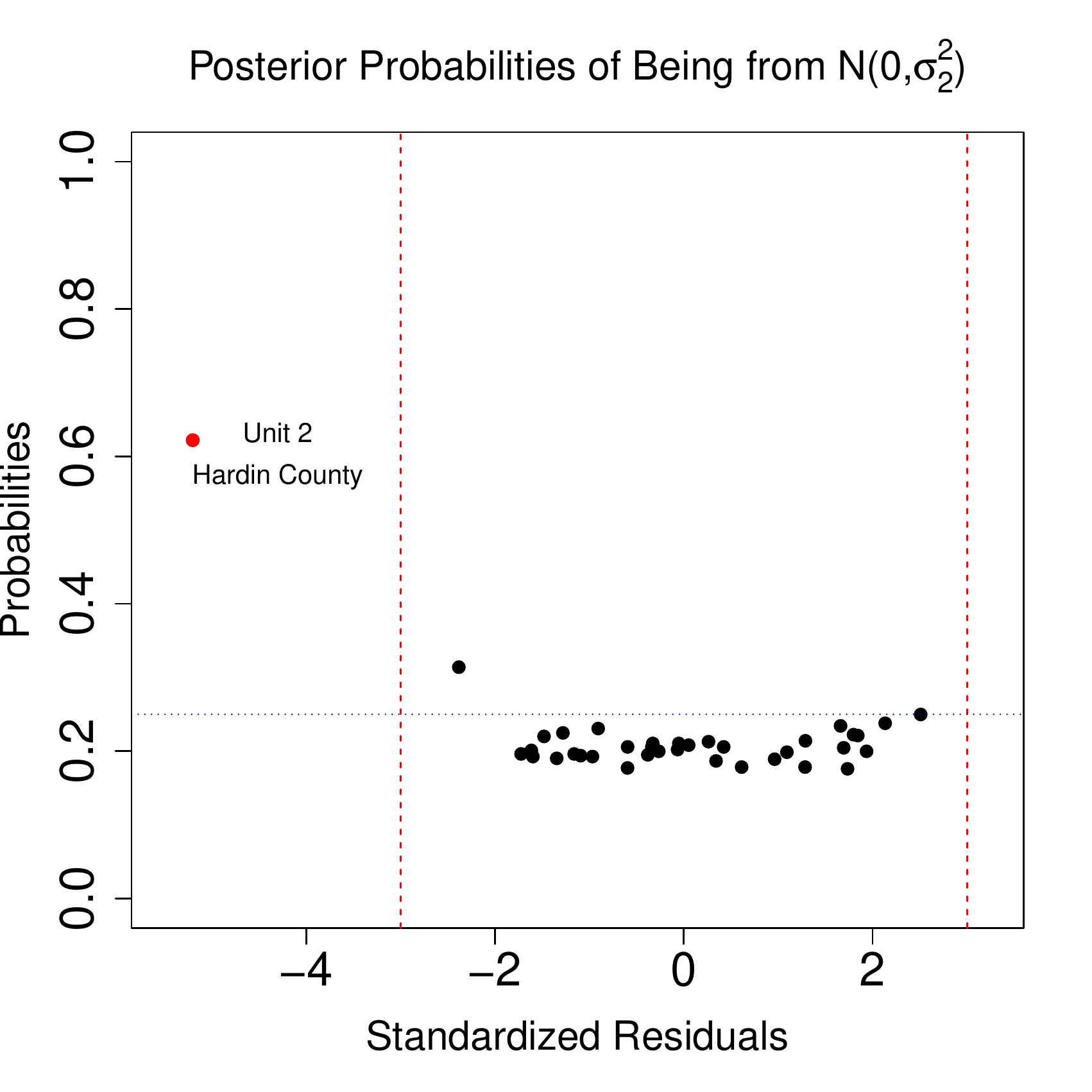} & \includegraphics[scale=.35]{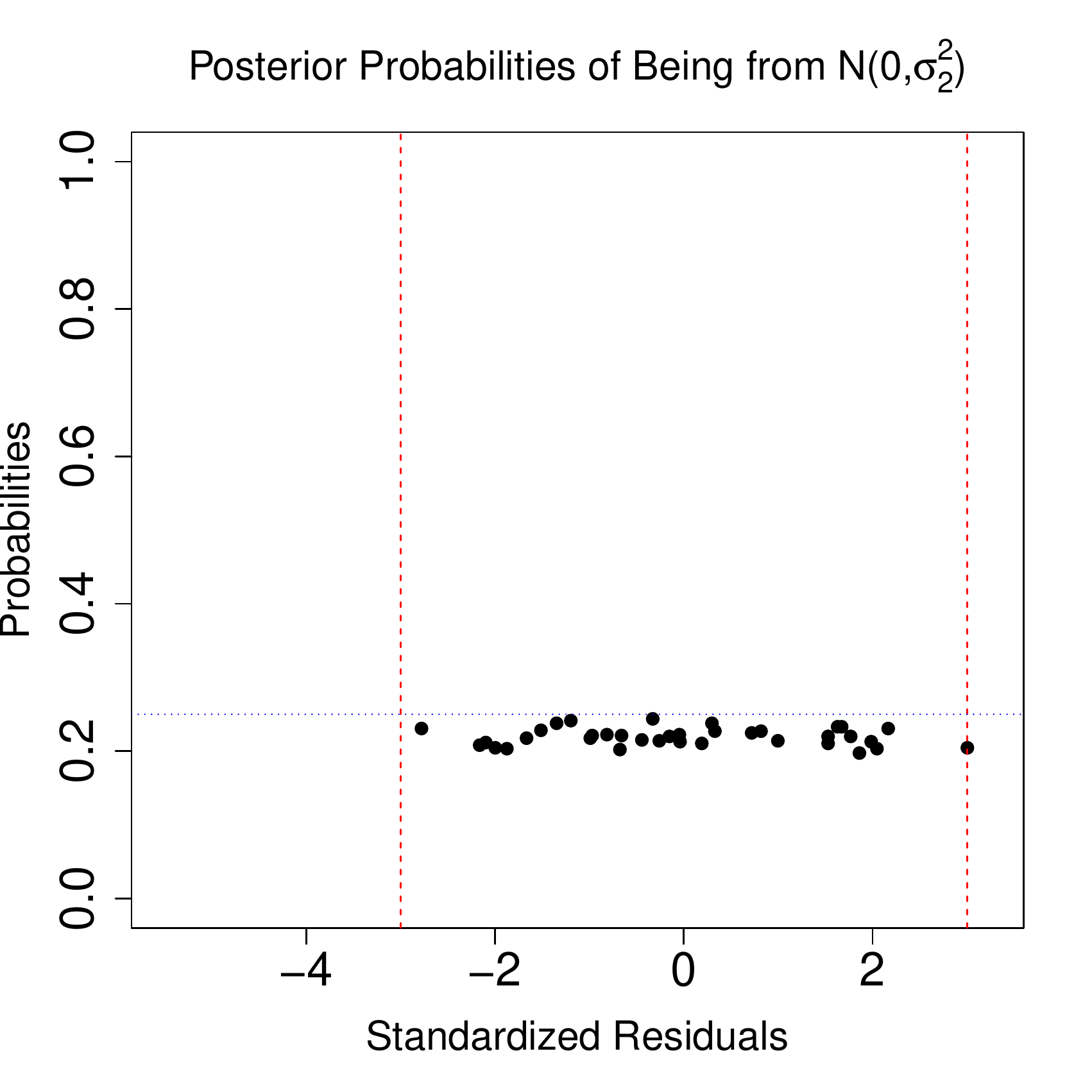} \\
			\end{tabular}
			\caption{Posterior probabilities of observations coming from subpopulation 2 in {\it full} and {\it reduced} corn data }\label{BHF-Reduced}
		\end{center}
	\end{figure}
	
	\noindent Tables 3 and 4 show posterior means and posterior medians, respectively, for $\beta_0, \beta_1, \beta_2, p_e, \sigma_v^2, \sigma_1^2,$ and $\sigma_2^2$. The estimated values of $\beta_0, \beta_1,$ and $\beta_2$ found from various methods appear to be similar, despite the difference in priors for $(\sigma_1^2, \sigma_2^2)$ and $p_e$. We note that the estimate of $p_e$ is higher when using the proposed HB model, which constrains $p_e$ between \rt{$2^{-1}$} and $1$, than when using the CDM model, which does not constrain $p_e$ but constrains $\sigma_1^2<\sigma_2^2$. In the proposed method, we define the primary variance $\sigma_1^2$ as the variance of the distribution from which more than 50\% of observations occur and the secondary variance $\sigma_2^2$ for the distribution of the remaining observations. {When examining the full data, we calculate the posterior mean and median estimates of $\sigma_1^2$ to be 246.33 and 203.78 respectively, while those for $\sigma_2^2$ are 1059.20 and 533.24 respectively. We can compare these values to the estimates produced using the CDM HB approach, where the primary distribution is defined as the one with the smaller variance. Using the CDM method and the full set of data, we find the posterior mean and median estimates of $\sigma_1^2$ to be 186.95 and 173.04 respectively, and those of $\sigma_2^2$ as 842.25 and 480.48 respectively. Notably, in both methods, the primary population occurs with {$p_e>2^{-1}$} and has the smaller variance $\sigma_1^2$.}

	\begin{sidewaystable}
		\caption{Posterior means and standard deviations for relevant parameters in various models with and without the suspected outlier for corn data}\label{tabMPE:BHFanalysisMean}
		\begin{center}
			\begin{tabular}{c|rrrr|rrrr|rrrr}
				\hline
				& \multicolumn{4}{c}{DG} & \multicolumn{4}{c}{CDM} & \multicolumn{4}{c}{GDM}\\
				Estimates       & \multicolumn{2}{c}{Full Data} & \multicolumn{2}{c}{Reduced Data} & \multicolumn{2}{c}{Full Data} & \multicolumn{2}{c}{Reduced Data} & \multicolumn{2}{c}{Full Data} & \multicolumn{2}{c}{Reduced Data}\\
				& Mean     & SD       & Mean     & SD      & Mean     & SD        & Mean      & SD         & Mean      & SD        & Mean     & SD \\
				\hline                                                                                          
				$\hat{\beta}_0$      & $ 17.29$ & $ 36.48$ & $ 50.46$ & $28.43$ & $ 30.55$ & $  31.67$ & $  51.17$ & $   25.14$ & $ 33.26 $  & $30.98$     & $ 50.98$  & $27.28$    \\
				$\hat{\beta}_1$      & $  0.37$ & $  0.08$ & $  0.33$ & $ 0.06$ & $  0.35$ & $   0.06$ & $   0.33$ & $    0.05$ & $  0.35 $  & $0.06 $     & $  0.33$  & $0.06 $ \\
				$\hat{\beta}_2$      & $ -0.03$ & $  0.08$ & $ -0.13$ & $ 0.06$ & $ -0.07$ & $  0.08$ & $  -0.14$ & $    0.06$ & $ -0.08 $  & $0.07 $     & $ -0.14$  & $0.06 $    \\
				$\hat{p}_e$          & $     -$ & $     -$ & $     -$ & $    -$ & $  0.60$ & $   0.27$ & $   0.47$ & $    0.29$ & $  0.77 $  & $0.14 $     & $  0.78$  & $0.15 $    \\
				$\hat{\sigma}_v^2$   & $175.08$ & $193.42$ & $228.88$ & $178.87$& $203.53$ & $ 178.55$ & $ 257.32$ & $  183.64$ & $ 205.71$  & $176.27$    & $266.17$  & $202.47$    \\
				$\hat{\sigma}_1^2$   & $364.47$ & $144.71$ & $210.48$ & $89.81$ & $186.95$ & $ 100.36$ & $ 118.96$ & $   56.58$ & $ 246.33$  & $158.54$    & $166.58$  & $68.51$    \\
				$\hat{\sigma}_2^2$   & $     -$ & $     -$ & $     -$ & $    -$ & $842.25$ & $2090.03$ & $ 364.21$ & $ 1580.03$ & $1059.20$  & $2487.99$   & $233.30$  & $432.57$    \\
				\hline                                                                                                                         
			\end{tabular}		
		\end{center}
		\vspace{2\baselineskip}
		
		\caption{Posterior medians and interquartile ranges for relevant parameters in various models with and without the suspected outlier for corn data}\label{tabMPE:BHFanalysisMedian}
		\begin{center}
			\begin{tabular}{c|rrrr|rrrr|rrrr}
				\hline
				& \multicolumn{4}{c}{DG} & \multicolumn{4}{c}{CDM} & \multicolumn{4}{c}{GDM}\\
				Estimates       & \multicolumn{2}{c}{Full Data} & \multicolumn{2}{c}{Reduced Data} & \multicolumn{2}{c}{Full Data} & \multicolumn{2}{c}{Reduced Data} & \multicolumn{2}{c}{Full Data} & \multicolumn{2}{c}{Reduced Data}\\
				& Median   & IQR      & Median   & IQR      & Median   & IQR       & Median    & IQR        & Median     & IQR           & Median    & IQR \\
				\hline                                                                                          
				$\hat{\beta}_0$      & $ 18.87$ & $ 43.37$ & $ 50.22$ & $ 36.96$ & $ 30.59$ & $  42.12$ & $  53.40$ & $  31.34$  & $ 34.03 $  & $  42.54$ & $ 51.19$  & $34.78$    \\
				$\hat{\beta}_1$      & $  0.37$ & $  0.10$ & $  0.33$ & $  0.07$ & $  0.35$ & $   0.09$ & $   0.33$ & $   0.06$  & $  0.35 $  & $  0.08 $ & $  0.33$  & $0.07 $ \\
				$\hat{\beta}_2$      & $ -0.03$ & $  0.10$ & $ -0.13$ & $  0.08$ & $ -0.07$ & $   0.10$ & $  -0.14$ & $   0.08$  & $ -0.08 $  & $  0.09 $ & $ -0.14$  & $0.08 $    \\
				$\hat{p}_e$          & $     -$ & $     -$ & $     -$ & $     -$ & $  0.66$ & $   0.44$ & $   0.45$ & $   0.53$  & $  0.79 $  & $  0.24 $ & $  0.79$  & $0.25 $    \\
				$\hat{\sigma}_v^2$   & $123.56$ & $153.47$ & $188.54$ & $169.17$ & $157.17$ & $ 170.35$ & $ 208.56$ & $ 183.63$  & $ 151.16$  & $ 171.94$ & $206.04$  & $199.12$ \\
				$\hat{\sigma}_1^2$   & $334.70$ & $152.13$ & $191.44$ & $ 86.37$ & $173.04$ & $ 120.84$ & $ 114.18$ & $  65.50$  & $ 203.78$  & $ 180.15$ & $155.76$  & $72.71$    \\
				$\hat{\sigma}_2^2$   & $     -$ & $     -$ & $     -$ & $     -$ & $480.28$ & $ 447.43$ & $ 191.71$ & $  93.85$  & $ 533.24$  & $ 852.00$ & $136.73$  & $165.04$    \\
				\hline                                                                                                                         
			\end{tabular}		
		\end{center}
	\end{sidewaystable}

	\subsection{AAGIS Farm Data Analysis}
	Chambers et al. (2011) considered data from the Australian Agricultural and Grazing Industries Survey (AAGIS) to provide at the regional level the estimated Total Cash Costs (TCC) associated with operation of a farm based on the farm area covariate. In our illustration we treated their sampled data of 1,652 farms as the {\it finite} population with 27 small areas. In the original dataset, there were 29 small areas. We merged two small areas which had small values of $N_i$ with the neighboring ones. From this population we considered a random sample of 50 units to create our working sample.  We drew a sample of 50 units with probabilities proportional to the sizes of the small areas. These 50 data points, along with the identification codes of the 27 small areas are given in Table 5. Here the response $Y$ is the total cash costs associated with operation of the farms, and we consider the farm area as the predictor variable $x$. A preliminary analysis of the data indicated a long right-tail for the response. To address this excessive skewness, we consider a logarithm transformation of the original response. We also use a similar transformation for the covariate $x$, the farm area.

	\begin{table}\caption{Small Areas and Samples from the AAGIS Farm Data}\label{Tab:Sampled_Data}
		\begin{minipage}[t]{0.3\textwidth}
			\bigskip
			
			{\footnotesize
				\begin{center}
					\begin{tabular}{lrr}
						\hline
						Small && \\
						Area   &      $y_{ij}$  &      $X_{ij}$    \\
						\hline
						111 &    453006 &   48583.0   \\
						121 &    144606 &     647.7   \\
						121 &   1212066 &   11660.0   \\
						121 &  16695291* &     445.5   \\
						122 &    140520 &    1042.0   \\
						122 &    137756 &    2063.9   \\
						122 &    198754 &    1978.0   \\
						123 &     83055 &     628.0   \\
						123 &    245025 &    1205.3   \\
						123 &    106124 &     491.0   \\
						131 &    167385 &    1021.0   \\
						131 &    335802 &    1807.0   \\
						132 &    134251 &    2332.0   \\
						221 &     47380 &     652.3   \\
						221 &    231261 &    2630.0   \\
						222 &     68023 &     683.8   \\ 
						222 &     60066 &    1881.0   \\
						\hline
					\end{tabular}
				\end{center}
			} 
		\end{minipage}
		\begin{minipage}[t]{0.3\textwidth}
			\bigskip
			
			{\footnotesize
				\begin{center}
					\begin{tabular}{lrr}
						
						\hline
						Small && \\
						Area   &      $y_{ij}$  &      $X_{ij}$    \\
						\hline
						223 &  31913070* &     260.1   \\
						223 &     18592 &      40.5   \\
						231 &    108257 &     744.7   \\
						231 &    145922 &     279.0   \\
						312 &    410995 &   48526.0   \\
						313 &     21792 &    3200.0   \\
						314 &    307842 &   12040.0   \\
						321 &     50352 &    1251.0   \\
						321 &    140634 &    3989.0   \\
						322 &    149343 &    1537.9   \\
						322 &     38283 &    8461.5   \\
						322 &    188839 &    2443.3   \\
						322 &    254143 &    1603.0   \\
						331 &     96744 &    1862.0   \\
						331 &    269170 &   25101.2   \\ 
						332 &    216304 &   23083.9   \\
						\hline
					\end{tabular}
				\end{center}
				\vspace{.1in}  * suspected with high unit error variance from a subpopulation
			}
		\end{minipage}
		\begin{minipage}[t]{0.3\textwidth}
			\bigskip
			
			{\footnotesize
				\begin{center}
					\begin{tabular}{lrr}
						
						\hline
						Small && \\
						Area   &      $y_{ij}$  &      $X_{ij}$    \\
						\hline
						411 &     47169 &    2985.0   \\
						421 &     80999 &     838.0   \\
						421 &    121788 &    2886.6   \\
						422 &     63476 &     362.3   \\
						422 &     54554 &     288.0   \\
						431 &    123407 &    1135.7   \\
						431 &     55208 &     500.0   \\
						512 &    216138 &  176732.0   \\
						521 &    227858 &    2682.0   \\
						521 &    147555 &    1403.6   \\
						521 &     49280 &     354.1   \\
						522 &    157571 &    3152.3   \\
						531 &     82563 &     151.0   \\
						531 &    220028 &      40.0   \\
						631 &    599960 &    1126.4   \\
						631 &    263680 &     775.3   \\ 
						711 &    173869 &  120800.0   \\
						\hline
					\end{tabular}
				\end{center}
			}
		\end{minipage}
	\end{table}

	Following Equation (1), we fit a model $Y^*_{ij}=\beta_0 + \beta_1x^*_{ij}+v_i+e_{ij}$ to predict the $m=27$ small area means  $\theta^*_i=\beta_0+\beta_1\bar{x}^*_{i(p)}+v_i$ of $Y_{ij}$'s, for $i=1,\ldots, 27$, where ${x}^*_{ij}=\log{({x}_{ij})}$ and $Y^*_{ij} = \log(Y_{ij})$. \bt{We use the HB model to predict $\theta_i=\exp(\theta_i^*)$, as prediction in the original scale of the response is preferable. Here $\theta_i$ is unknown but the finite population is known, so we approximate $\theta_i$ by $\bar{Y}_{iG}=\big(\prod_{j=1}^{N_i}Y_{ij}\big)^{1/N_i}$, the geometric mean of all the responses of all the units in the $i$th small area.}

	The predictors $\hat{\theta}_i$'s are calculated for DG, CDM and GDM methods, and {compared with} $\bar{Y}_{iG}$'s. Since the posterior distributions are long-tailed (to the right), we use the median of the $\hat{\theta}_{i,k}$ values, given by $\exp(\beta_{0,k} + \beta_{1,k}\bar{x}^*_{i} + v_{i,k})$, to estimate ${\theta}_{i}$.  To evaluate the effectiveness of an estimator {$\hat{\theta}_i$,} we computed the following four deviation measures for the estimator from the {``truth'';} the average absolute deviation (AAD), the average squared deviation (ASD), average absolute relative deviation (AARD) and the average squared  relative deviation (ASRD). 
	\begin{eqnarray*}
		\mbox{AAD}(\hat{\theta}) = \frac{1}{m}\sum_{i=1}^{m} |\hat{\theta}_i - \bar{Y}_{iG}|,~~&& ~~
		\mbox{ASD}(\hat{\theta}) = \frac{1}{m}\sum_{i=1}^{m} (\hat{\theta}_i - \bar{Y}_{iG})^2,\\
		\mbox{AARD}(\hat{\theta}) = \frac{1}{m}\sum_{i=1}^{m} \frac{|\hat{\theta}_i - \bar{Y}_{iG}|}{\bar{Y}_{iG}},~~&&~~
		\mbox{ASRD}(\hat{\theta}) = \frac{1}{m}\sum_{i=1}^{m} \frac{(\hat{\theta}_i - \bar{Y}_{iG})^2}{\bar{Y}_{iG}^2}.\\
	\end{eqnarray*}  
	These summary measures for the three competing methods are given in Table 6.

	\begin{table}[h]\caption{Performance of competing methods}\label{Tab1:performance}
		\bigskip
		\begin{center}
			\begin{tabular}{ll|cccc}
				\hline
				&&      AAD    &        ASD  &    AARD     &   ASRD   \\
				\hline
				& DG & 50168 & 4865362824 & 0.37 & 0.34 \\
				&CDM & 49059 & 4413325890 & 0.38 & 0.36 \\
				&GDM & 36857 & 2592492269 & 0.22 & 0.09 \\
				\hline
			\end{tabular}
		\end{center}
		
	\end{table}

	We also calculated 90\% credible intervals (CrI) for $\theta_i$ under the DG, CDM and GDM methods, and reported the ratios of their lengths in Table 7. In Figure 2 we plotted the posterior probabilities of each sampled {observation} coming from {the} subpopulation 2. We notice that the GDM method {correctly} identifies the {observations that are believed to have unit-level error distribution from subpopulation 2.}

	\begin{sidewaystable}
		\caption{Summary results of full AAGIS data analysis under GDM, CDM and DG methods }\label{Tab:Summary_results}
		\vspace{.2in}
		
		{\footnotesize
			\begin{center}
				\begin{tabular}{|lrrr|rrr|rr|rr|rr|rr|}
					\hline
					Small  &        &              & & \multicolumn{3}{c|}{Median Estimate} & \multicolumn{2}{c|}{90\% DG CrI} &  \multicolumn{2}{c|}{90\% CDM CrI} & \multicolumn{2}{c|}{90\% GDM CrI} & \multicolumn{2}{c|}{Lengths of CrIs} \\
					Area$_i$   &  $N_i$  & \multicolumn{1}{c}{$\bar{\bm{x}}^{*}_i$} & \multicolumn{1}{c|}{{$\bar{Y}_{iG}$}} &  $\hat{\theta}_{i,DG}$  & $\hat{\theta}_{i,CDM}$  & $\hat{\theta}_{i,GDM}$  & Lower  & Upper  & Lower & Upper & Lower & Upper & $\frac{DG}{GDM}$ & $\frac{CDM}{GDM}$ \\[10pt]
					\hline
					111 & 30 & 9.89 & 201370 & 226960 & 234020 & 274489 & 89632 & 659957 & 97726 & 636638 & 148199 & 510486 & 1.57 & 1.49 \\ 
					121 & 95 & 7.55 & 185680 & 299834 & 288577 & 207637 & 136725 & 995215 & 139375 & 883369 & 111519 & 424164 & 2.75 & 2.38 \\ 
					122 & 103 & 7.06 & 129304 & 153431 & 155675 & 134176 & 72126 & 328112 & 73394 & 308858 & 88888 & 201792 & 2.27 & 2.09 \\ 
					123 & 108 & 7.02 & 161197 & 148127 & 150148 & 133846 & 65722 & 318273 & 71636 & 303909 & 86265 & 203250 & 2.16 & 1.99 \\ 
					131 & 81 & 6.83 & 122631 & 161584 & 164871 & 155646 & 73644 & 388526 & 79466 & 378048 & 94019 & 266281 & 1.83 & 1.73 \\ 
					132 & 34 & 5.93 & 43616 & 133227 & 133394 & 87551 & 49643 & 318228 & 49162 & 308223 & 46084 & 156649 & 2.43 & 2.34 \\ 
					221 & 55 & 6.93 & 108188 & 138689 & 146352 & 108314 & 58761 & 312844 & 62624 & 304393 & 62320 & 186072 & 2.05 & 1.95 \\ 
					222 & 60 & 6.74 & 100614 & 123660 & 123288 & 80202 & 48018 & 268881 & 49547 & 243273 & 45724 & 136427 & 2.44 & 2.14 \\ 
					223 & 73 & 6.23 & 80062 & 207196 & 220491 & 76694 & 97501 & 676351 & 103365 & 642831 & 36630 & 156016 & 4.85 & 4.52 \\ 
					231 & 77 & 6.13 & 87463 & 133833 & 141195 & 109808 & 56316 & 318677 & 61254 & 304479 & 68852 & 180173 & 2.36 & 2.18 \\ 
					312 & 46 & 11.43 & 327596 & 264692 & 285847 & 400303 & 87422 & 905536 & 99404 & 854117 & 198483 & 812195 & 1.33 & 1.23 \\ 
					313 & 30 & 10.02 & 218926 & 161457 & 179806 & 180586 & 50375 & 408233 & 53825 & 435024 & 53637 & 473588 & 0.85 & 0.91 \\ 
					314 & 40 & 9.94 & 255903 & 218557 & 228872 & 279868 & 86875 & 642940 & 91385 & 600219 & 152722 & 546724 & 1.41 & 1.29 \\ 
					321 & 79 & 6.81 & 103095 & 129885 & 133642 & 85908 & 50192 & 281050 & 52393 & 272303 & 49185 & 146315 & 2.38 & 2.26 \\ 
					322 & 117 & 8.18 & 198718 & 161917 & 160211 & 182018 & 73386 & 332377 & 78158 & 307968 & 107870 & 296511 & 1.37 & 1.22 \\ 
					331 & 51 & 7.31 & 87874 & 153162 & 153655 & 115249 & 65710 & 358745 & 67244 & 336329 & 69382 & 189235 & 2.44 & 2.25 \\ 
					332 & 19 & 8.61 & 178834 & 183179 & 187933 & 161215 & 77278 & 464113 & 79663 & 437318 & 90782 & 280135 & 2.04 & 1.89 \\ 
					411 & 36 & 10.81 & 206493 & 193958 & 204356 & 213142 & 64162 & 565956 & 65898 & 566532 & 97021 & 454788 & 1.40 & 1.40 \\ 
					421 & 51 & 7.64 & 135527 & 150222 & 151062 & 121699 & 61357 & 312654 & 66887 & 324839 & 74755 & 192536 & 2.13 & 2.19 \\ 
					422 & 80 & 6.87 & 95703 & 128930 & 127823 & 94119 & 48322 & 269358 & 52539 & 256305 & 56972 & 148261 & 2.42 & 2.23 \\ 
					431 & 74 & 6.81 & 120257 & 133921 & 134183 & 98098 & 52504 & 292267 & 54869 & 281025 & 59034 & 158215 & 2.42 & 2.28 \\ 
					512 & 31 & 12.29 & 256312 & 261533 & 280981 & 344572 & 75382 & 965835 & 83982 & 878019 & 170695 & 761121 & 1.51 & 1.34 \\ 
					521 & 83 & 7.60 & 215403 & 151586 & 155214 & 138008 & 69269 & 319690 & 72285 & 303267 & 91402 & 208303 & 2.14 & 1.98 \\ 
					522 & 47 & 8.19 & 245206 & 171282 & 176102 & 158791 & 72338 & 410603 & 74490 & 419638 & 91303 & 282574 & 1.77 & 1.80 \\ 
					531 & 60 & 6.53 & 124681 & 149993 & 151891 & 139569 & 64711 & 352785 & 70671 & 332599 & 78966 & 280841 & 1.43 & 1.30 \\ 
					631 & 62 & 6.62 & 133561 & 180222 & 182837 & 197633 & 84495 & 461879 & 86294 & 450867 & 96456 & 391833 & 1.28 & 1.23 \\ 
					711 & 30 & 12.47 & 503157 & 261186 & 284549 & 347653 & 72572 & 984729 & 84398 & 927491 & 165286 & 788502 & 1.46 & 1.35 \\ 
					\hline
				\end{tabular}
			\end{center}
		}
	\end{sidewaystable}

	\begin{figure}[h]
		\begin{center}
			\begin{tabular}{c}
				\includegraphics[scale=.5]{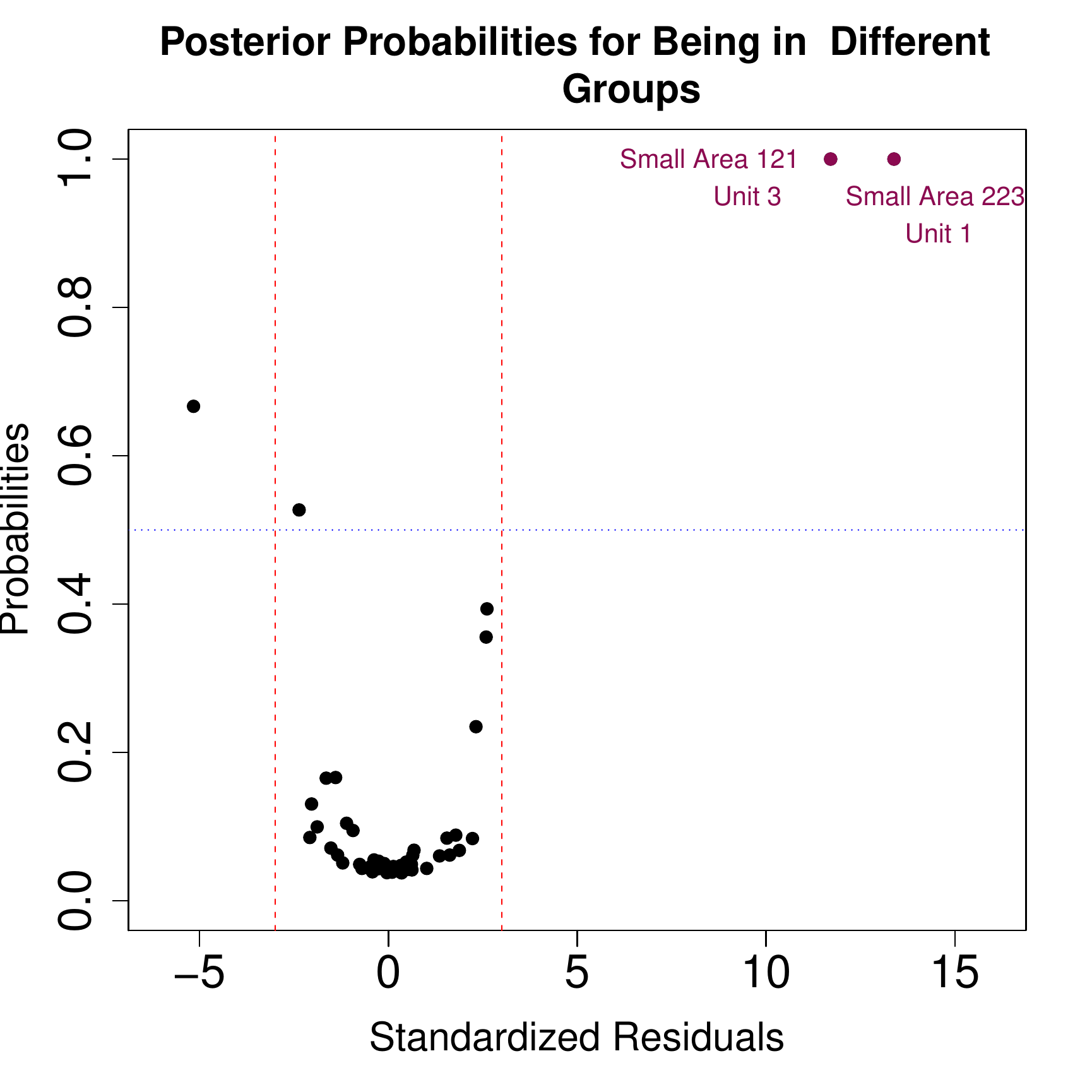}
			\end{tabular}
			\caption{Posterior probabilities of observations coming from subpopulation 2}\label{Fig:Post_Prob}
		\end{center}
	\end{figure}

	
	\section{Simulation Study}\label{sec:simul}
	Sinha and Rao (2009) and Chakraborty et al. (2018) employ a simulation study to evaluate and compare the performances of suggested extensions of NER models. We follow their example by first assuming a population with $m=40$ small areas, where each small area has $N_i=200$ units. We assume a single auxiliary variable $x_{ij}$ for each unit in the population, drawn independently from $N(1,1)$. The set of auxiliary variables $\mathbf{X}$ is kept fixed for all simulations.
	
	\bigskip\noindent For each simulation, we independently generate sets of area-level random effects $v_i$ for $i=1,\ldots,m$ from $N(0,1)$. Each small area has a population of size $N_i=200$. In the first four simulation setups, we generate $e_{ij}$ such that the mean of the unit-level errors is centered around 0. In these scenarios, we generate $e_{ij}$ from one of the four possible distributions: (i) all $e_{ij}$ are generated independently from $N(0,1)$; (ii) each $e_{ij}$ is drawn from $N(0, 1)$ with probability $p_e=0.90$ and from the secondary population with distribution $N(0,5^2)$ otherwise; (iii) each $e_{ij}$ is drawn from $N(0,1)$ with probability $p_e=0.60$ and from $N(0,5^2)$ otherwise; (iv) $e_{ij}$ are iid from a $t$-distribution with 4 degrees of freedom. We also perform a fifth simulation motivated by an example in Chambers et al. (2014) in which a very small portion of $e_{ij}$'s come from a secondary distribution with a non-zero mean. Here, each $e_{ij}$ is drawn from $N(0,1)$ with probability $p_e=0.97$ and from $N(5,5^2)$ otherwise. Setting $\beta_0 = 1$ and $\beta_1 = 1$ as in Sinha and Rao (2009) for each simulation method, we generate $m$ small area finite populations of $Y_{ij}=\beta_0 + \beta_1x_{ij}+v_i+e_{ij}$ based on Equation (1).

	\bigskip\noindent We compute a summary of auxiliary information for each small area as $\bar{X}_i=\frac{1}{N_i}\sum_{j=1}^{N_i}x_{ij}$ for $i=1,\ldots,m$. We then take a sample of size $n_i=4$ from each small area. Using auxiliary information, our goal is prediction of small area means $\bar Y_i =\frac 1{N_i}\sum_{j=1}^{N_i} Y_{ij}, i=1,\cdots, m$ for finite populations with large $N_i$ and small ratio $n_i/N_i$. From each sample, we derive HB predictors from the DG model and robust HB predictors from the outlier-accommodating CDM model and the more general proposed mixture model. These predictors are denoted as DG, CDM, and GDM, respectively, in subsequent data visualizations included in this paper. Since all three HB methods perform equally well when the unit-level errors contain no contamination, the plots for this simulation setup are relegated to Appendix A.2. We visualize the results of the other four simulation methods in Figures 3 to 5.

	\begin{figure}[h]
		\vspace{-.5cm}
		\begin{center}
			\begin{tabular}{ccc}
				\hspace{-.5in}\raisebox{-3cm}{40\% $N(0,5^2)$\hspace{.075cm}}     & \includegraphics[scale=.275,angle=270] {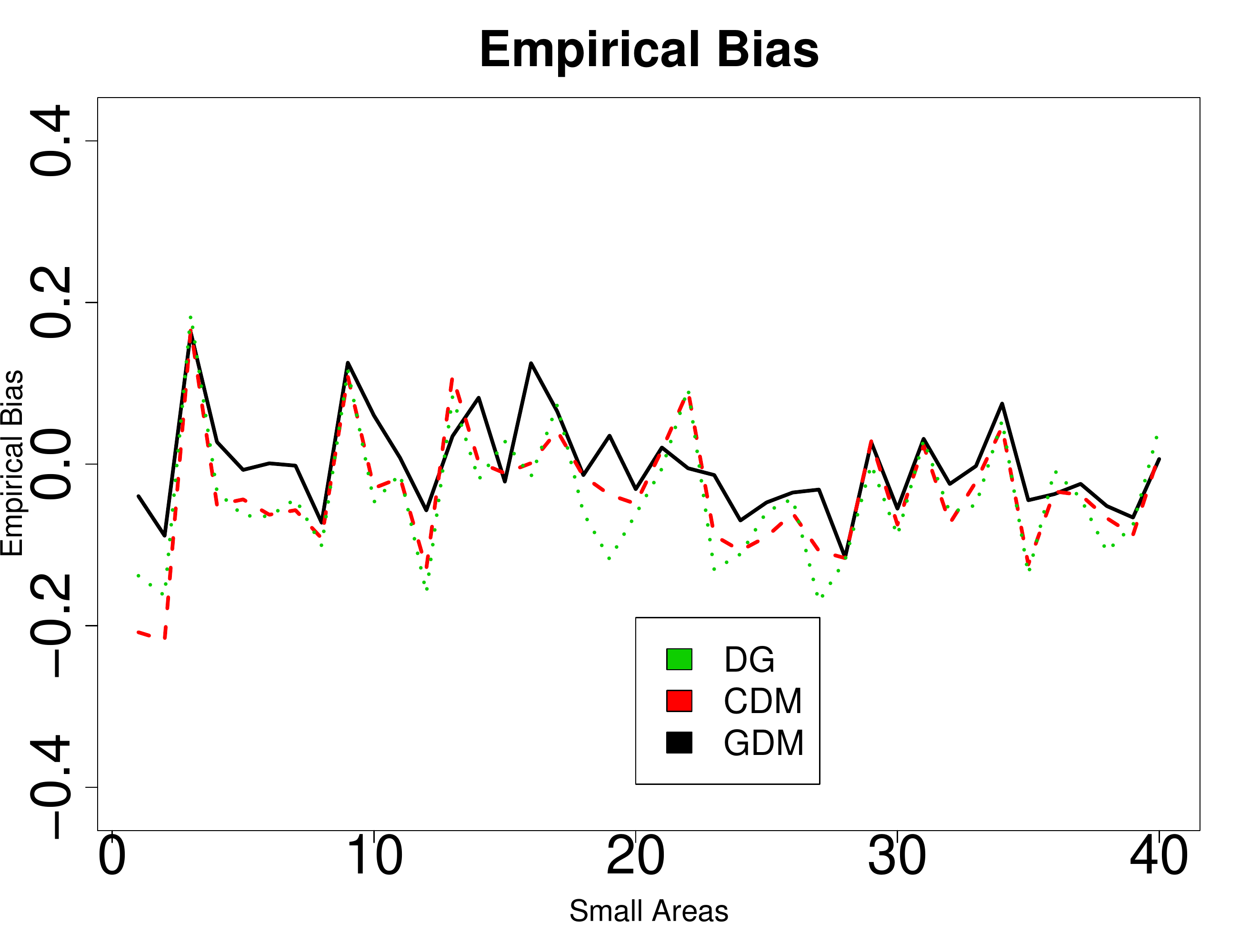} & \includegraphics[scale=.275,angle=270]{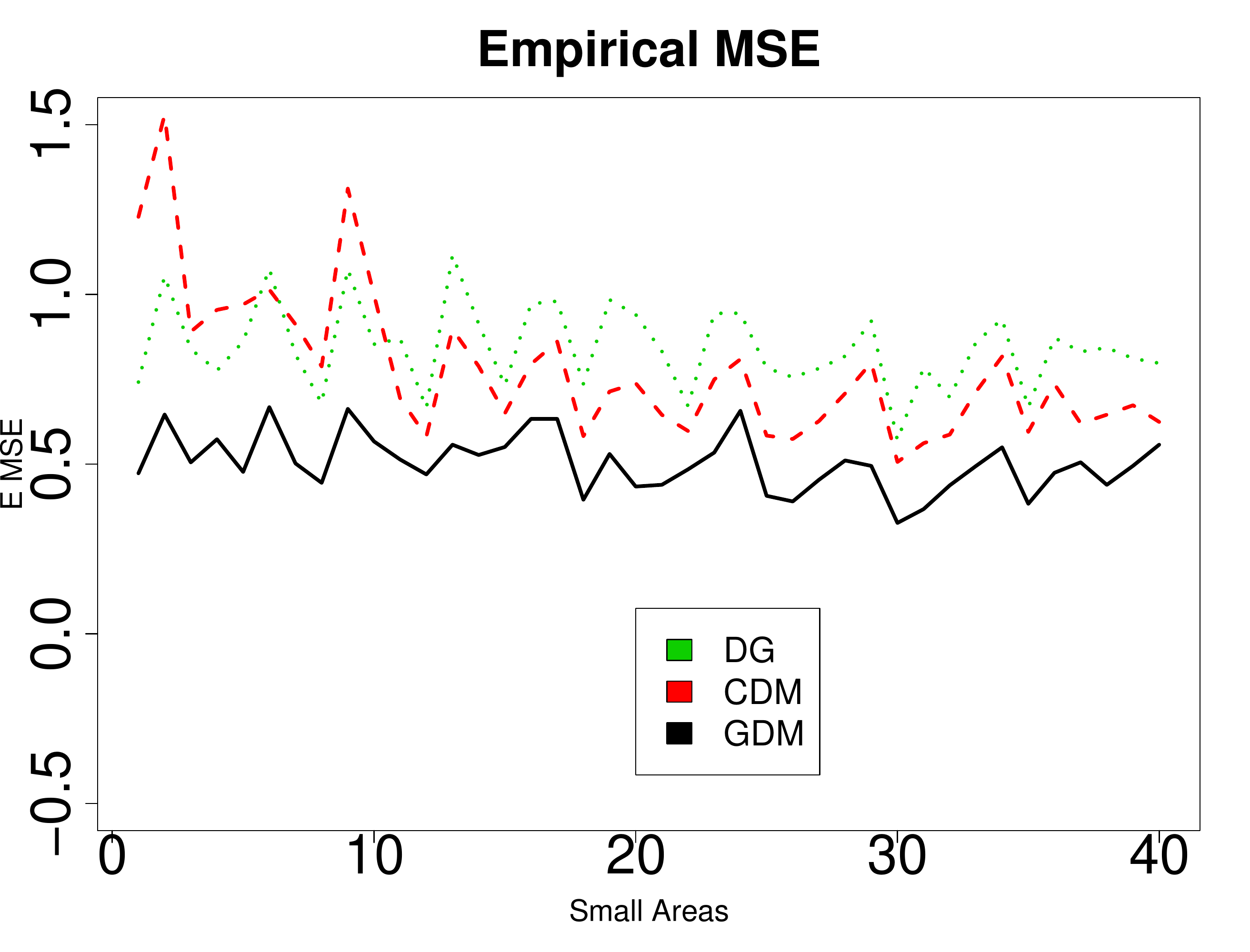}  \\
				\hspace{-.5in}\raisebox{-3cm}{10\% $N(0,5^2)$\hspace{.075cm}}     & \includegraphics[scale=.275,angle=270] {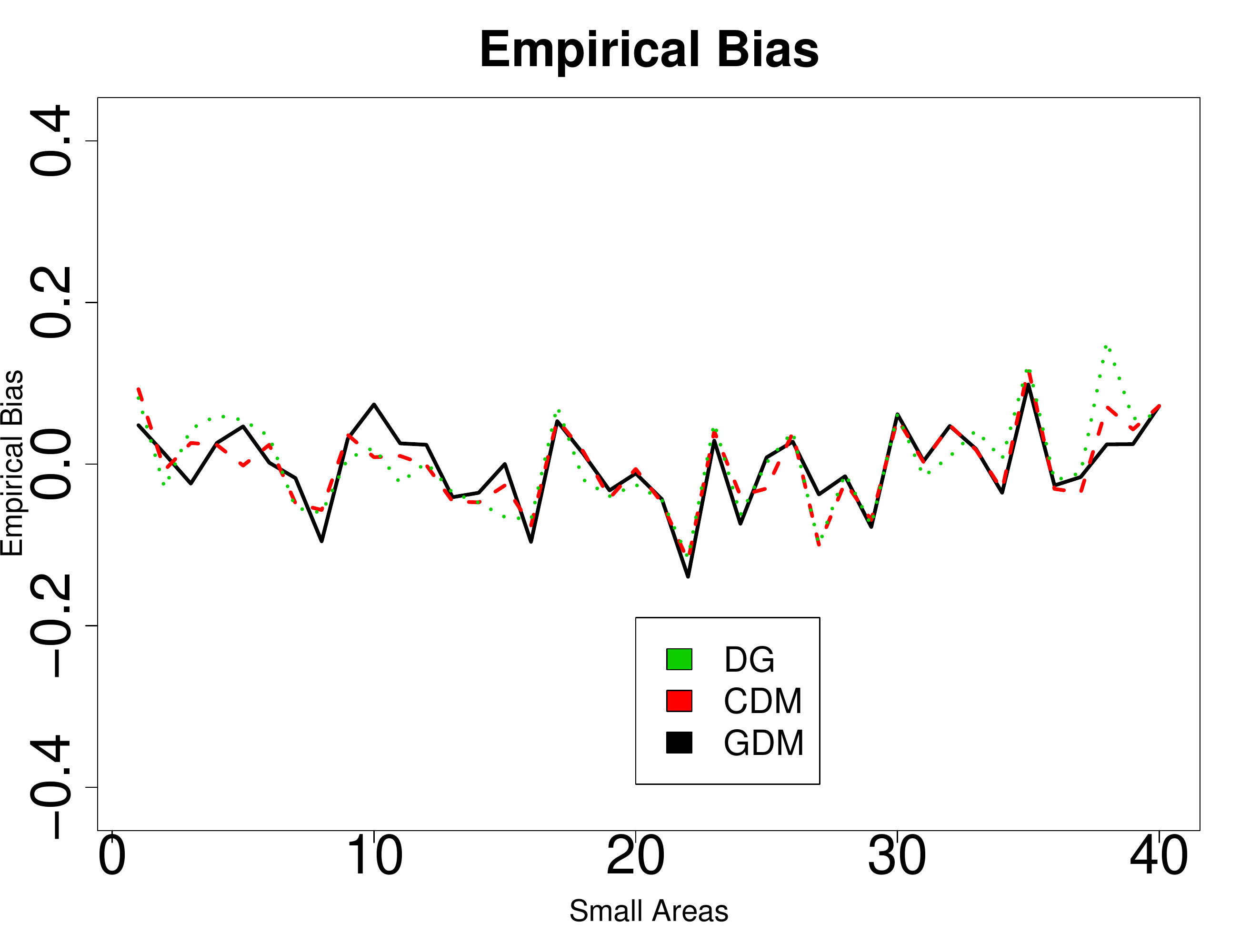} & \includegraphics[scale=.275,angle=270]{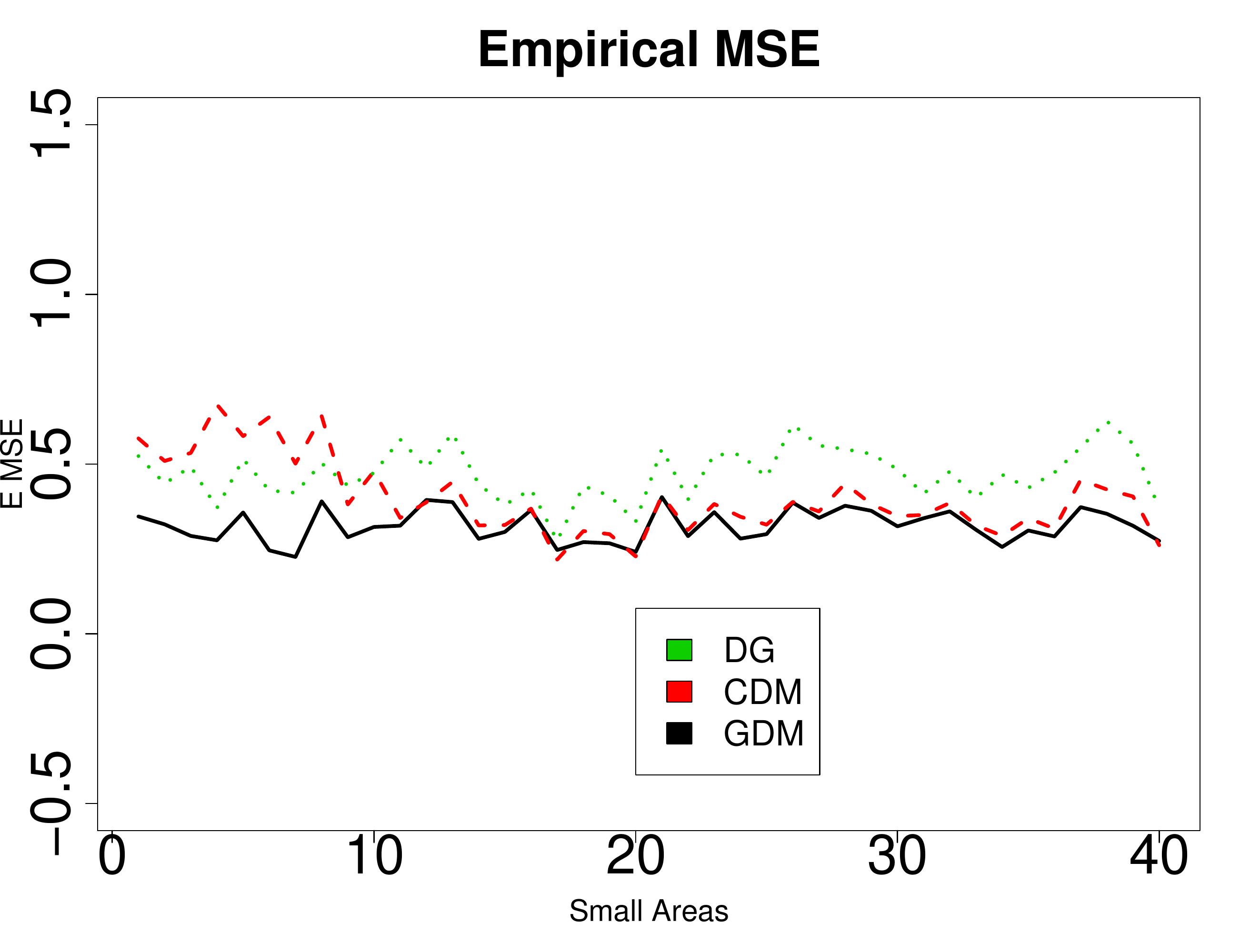}  \\
				\hspace{-.5in}\raisebox{-3cm}{3\% $N(5,5^2)$\hspace{.075cm}}     & \includegraphics[scale=.275,angle=270] {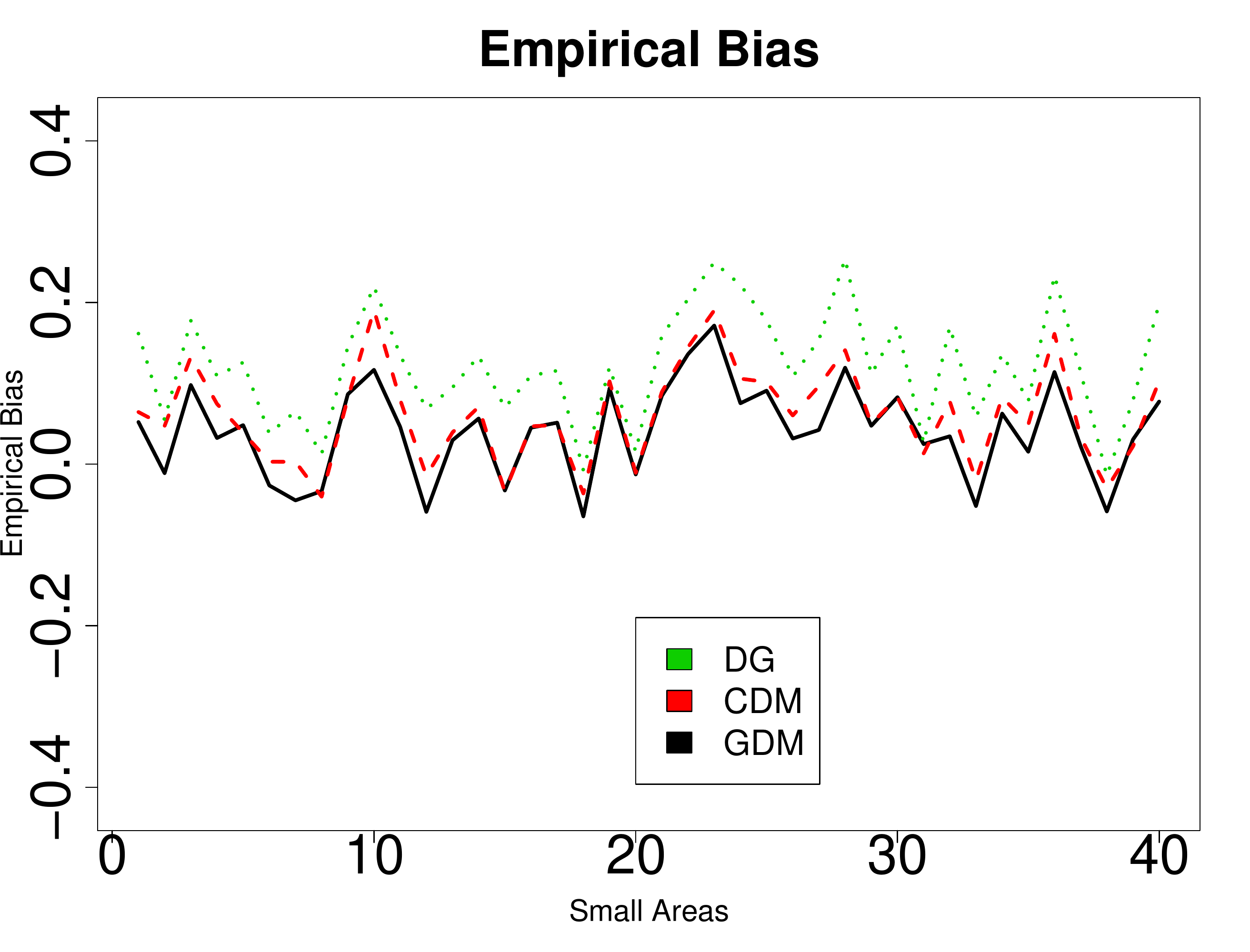} & \includegraphics[scale=.275,angle=270]{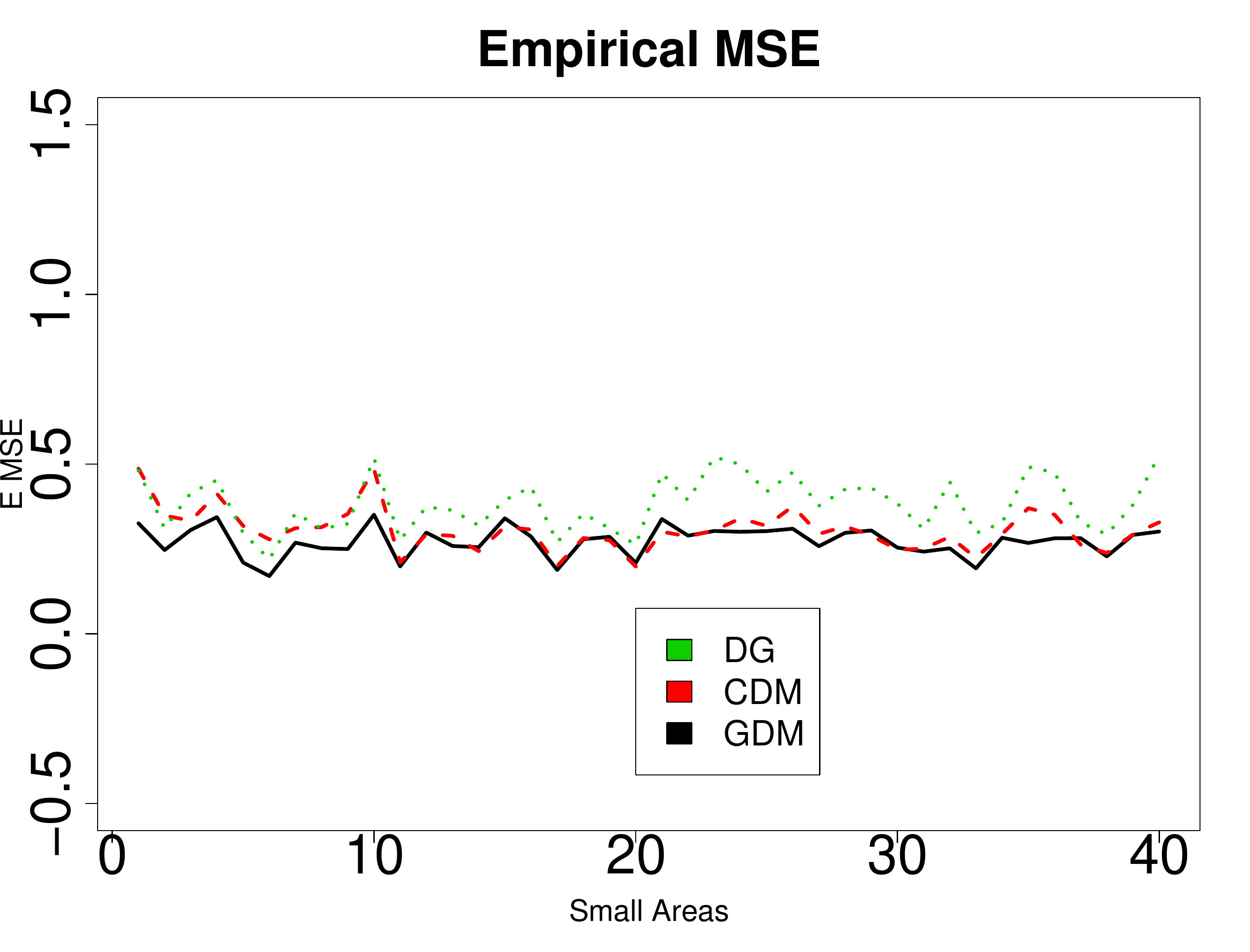}  \\
				\hspace{-.5in}\raisebox{-3cm}{$t_{(4)}$\hspace{.075cm}}     & \includegraphics[scale=.275,angle=270] {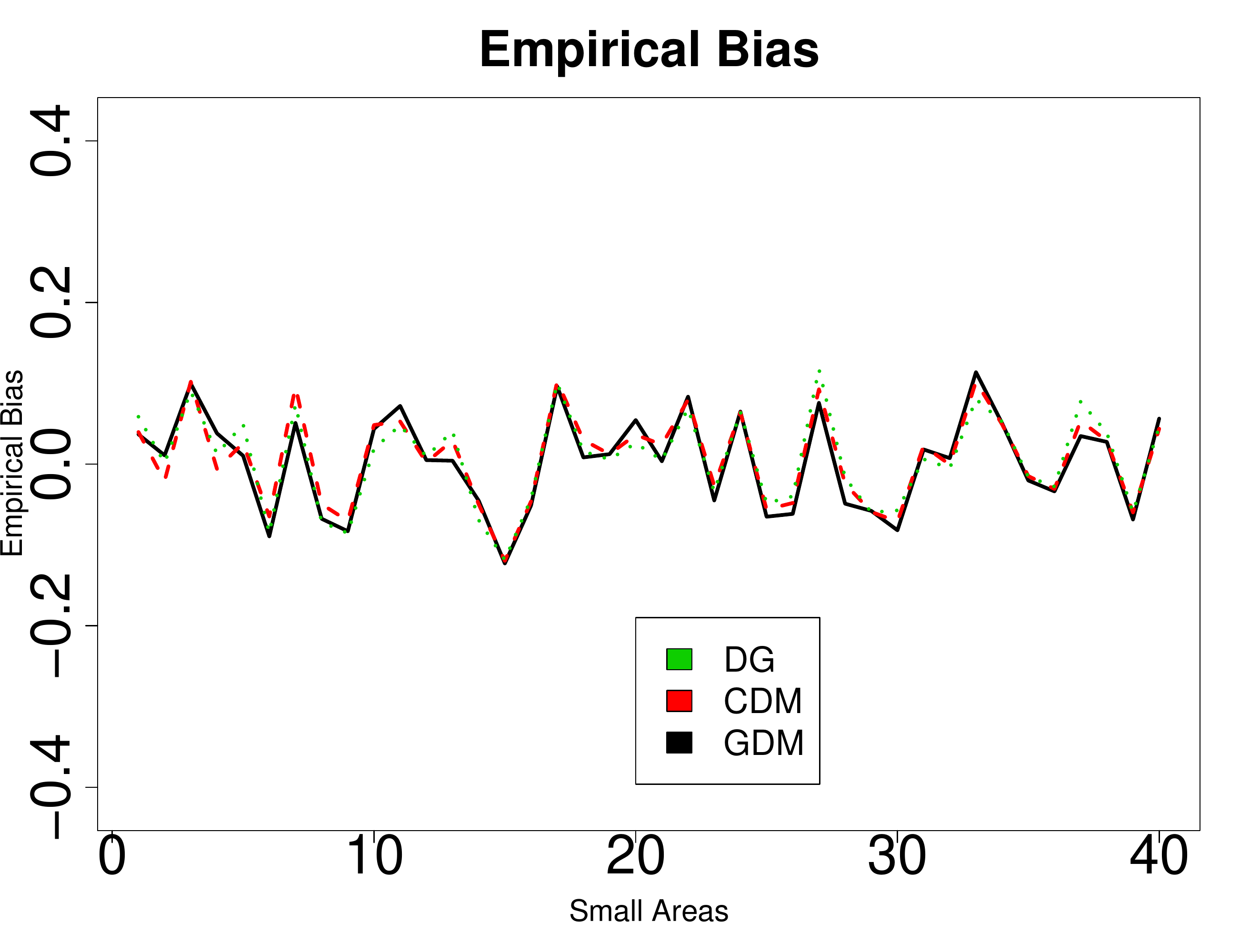} & \includegraphics[scale=.275,angle=270]{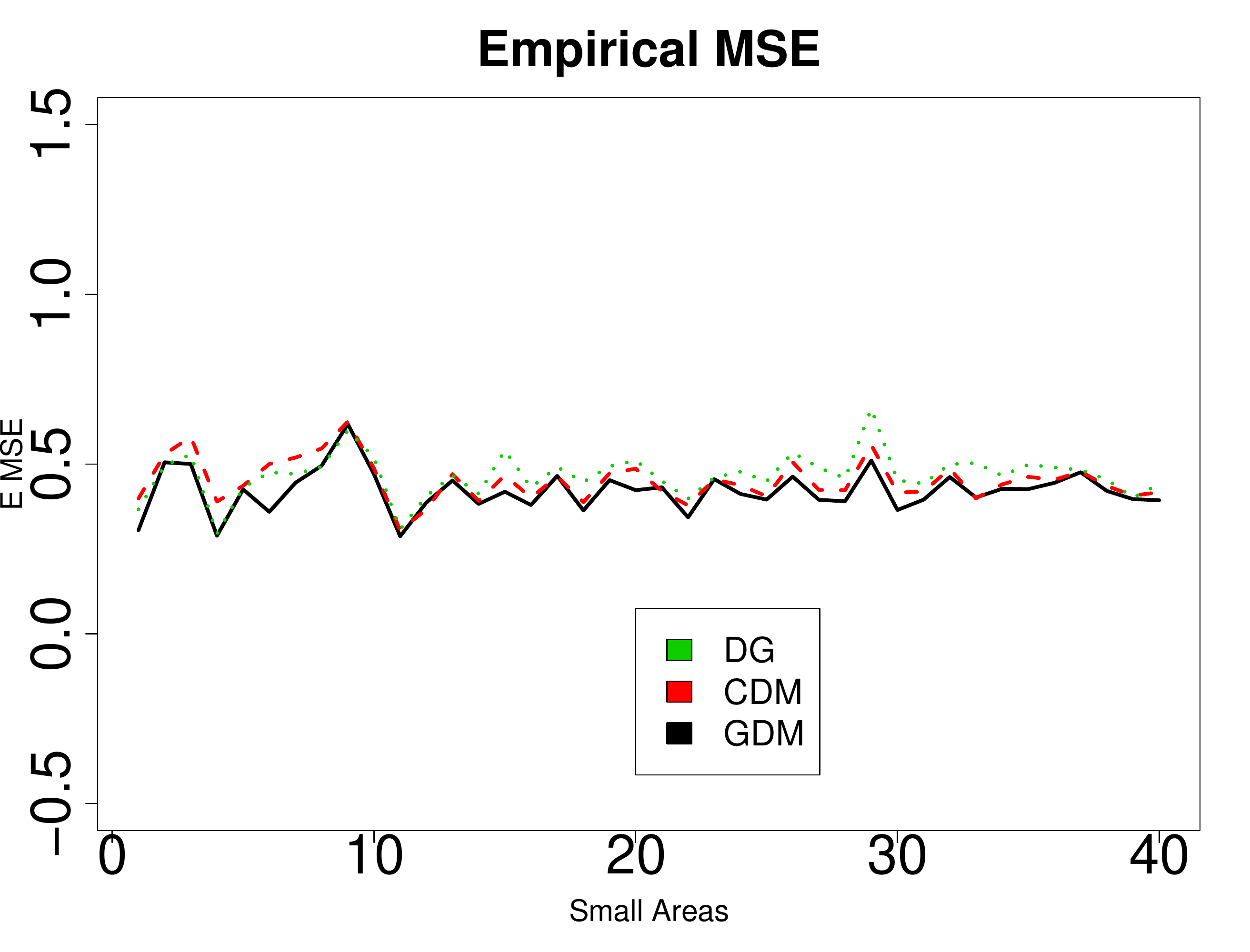}  \\
			\end{tabular}
			\caption{Plot of empirical biases and empirical MSEs of $\hat{\theta}$s}\label{tout:BiasVar3}
		\end{center}
	\end{figure}
	\clearpage
	
	\thispagestyle{empty}
	\begin{figure}[h]
		\vspace{-.5cm}
		\begin{center}
			\begin{tabular}{ccc}
				\hspace{-.5in}\raisebox{-3cm}{40\% $N(0,5^2)$\hspace{.075cm}}     & \includegraphics[scale=.275,angle=270] {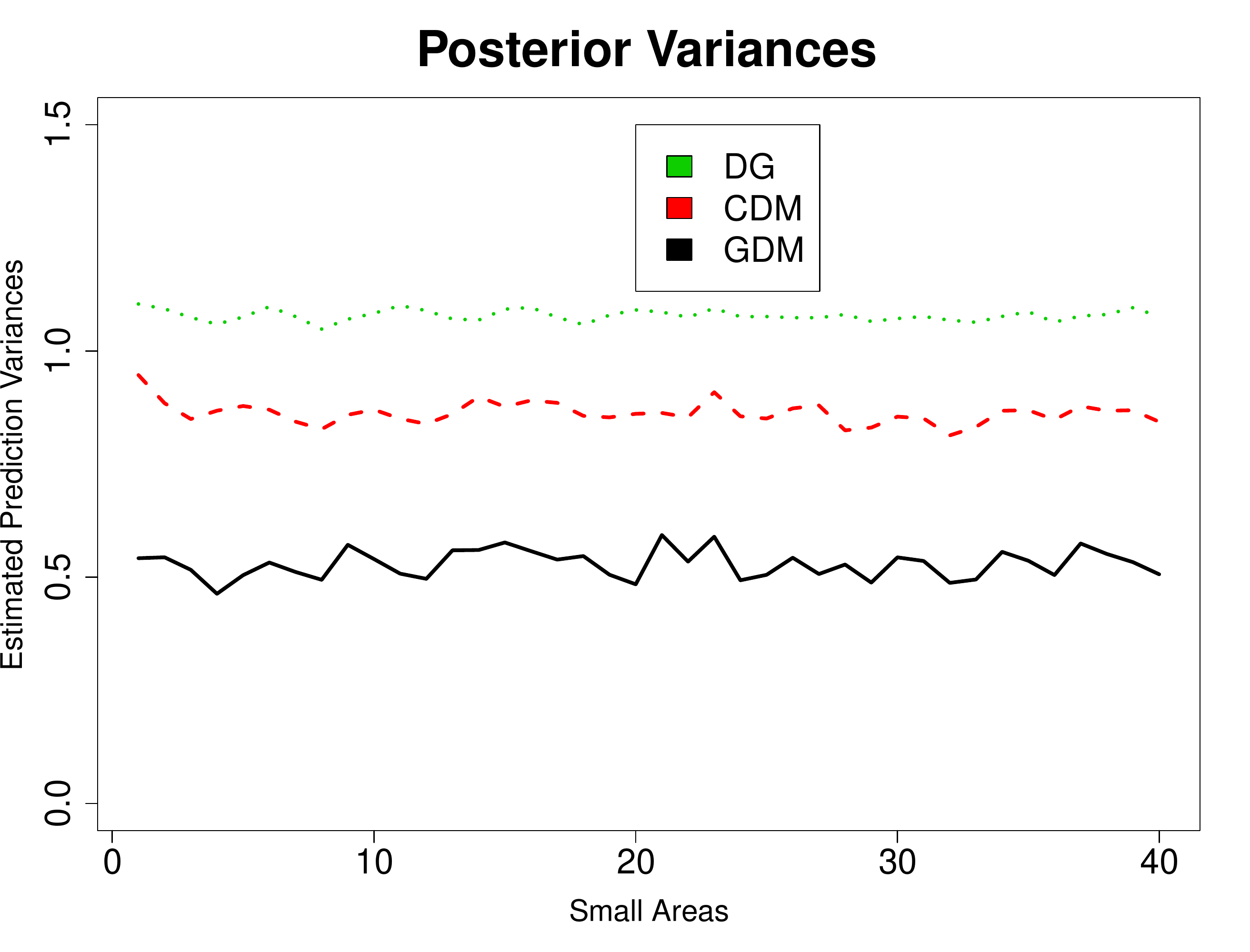} & \includegraphics[scale=.275,angle=270]{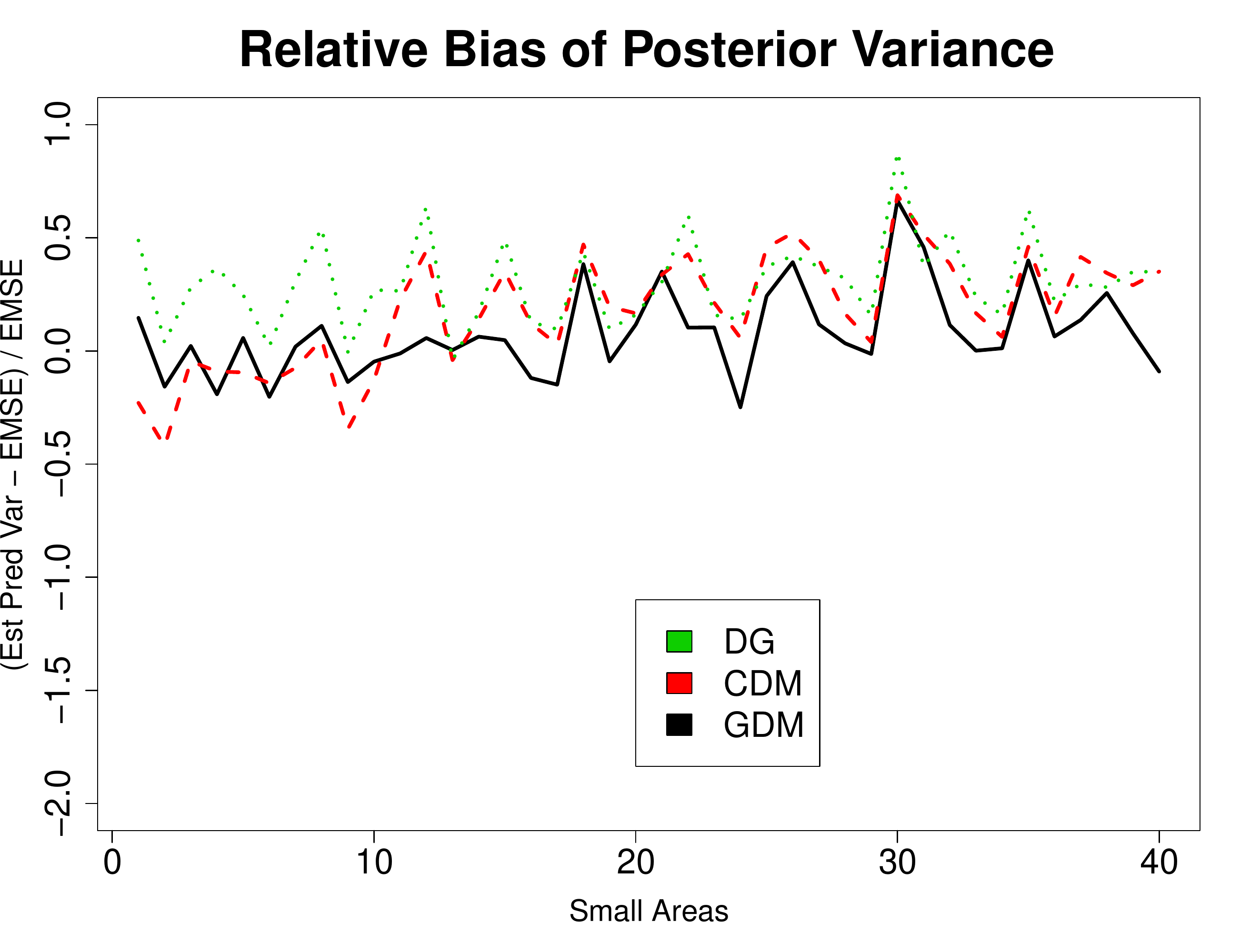}  \\
				\hspace{-.5in}\raisebox{-3cm}{10\% $N(0,5^2)$\hspace{.075cm}}     & \includegraphics[scale=.275,angle=270] {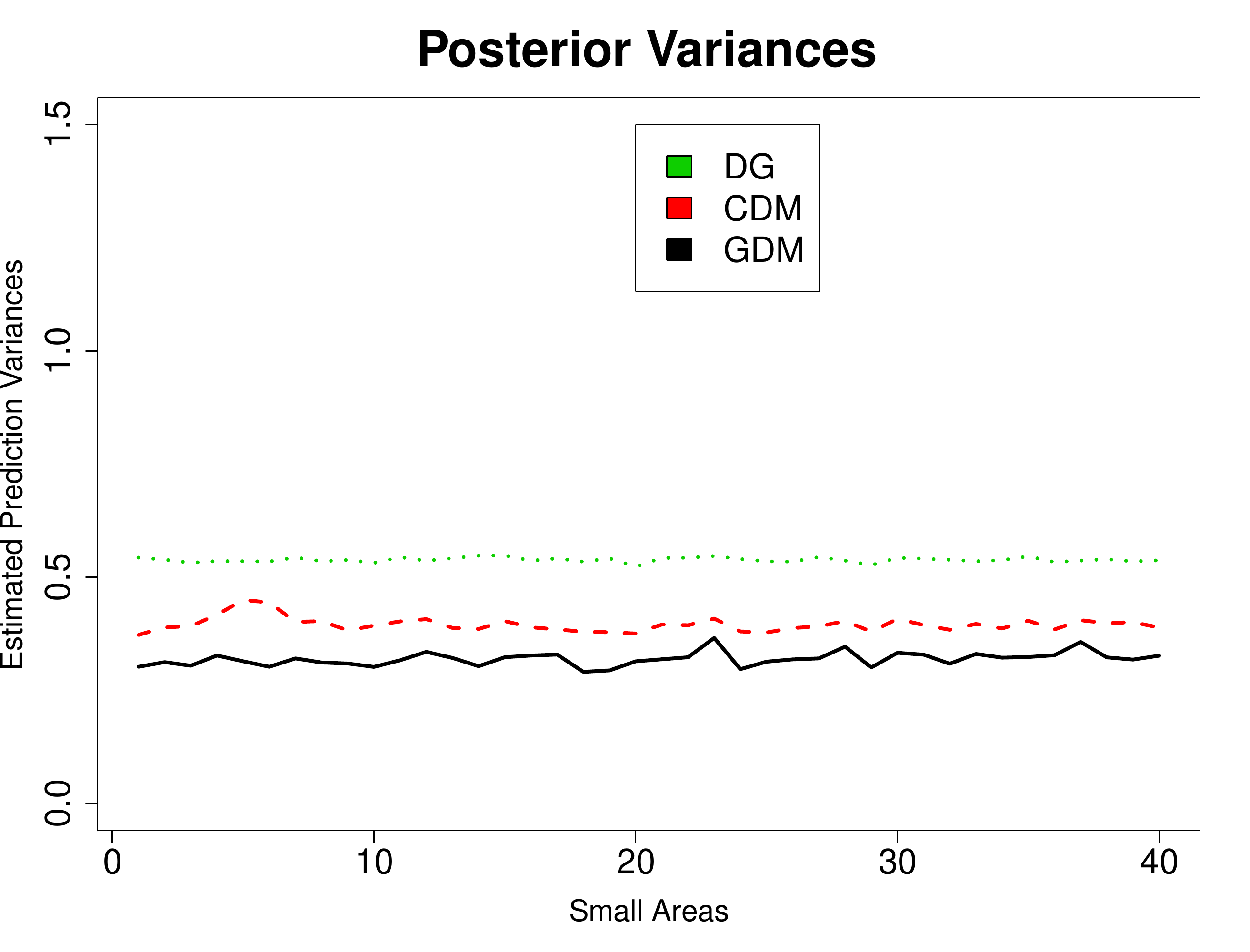} & \includegraphics[scale=.275,angle=270]{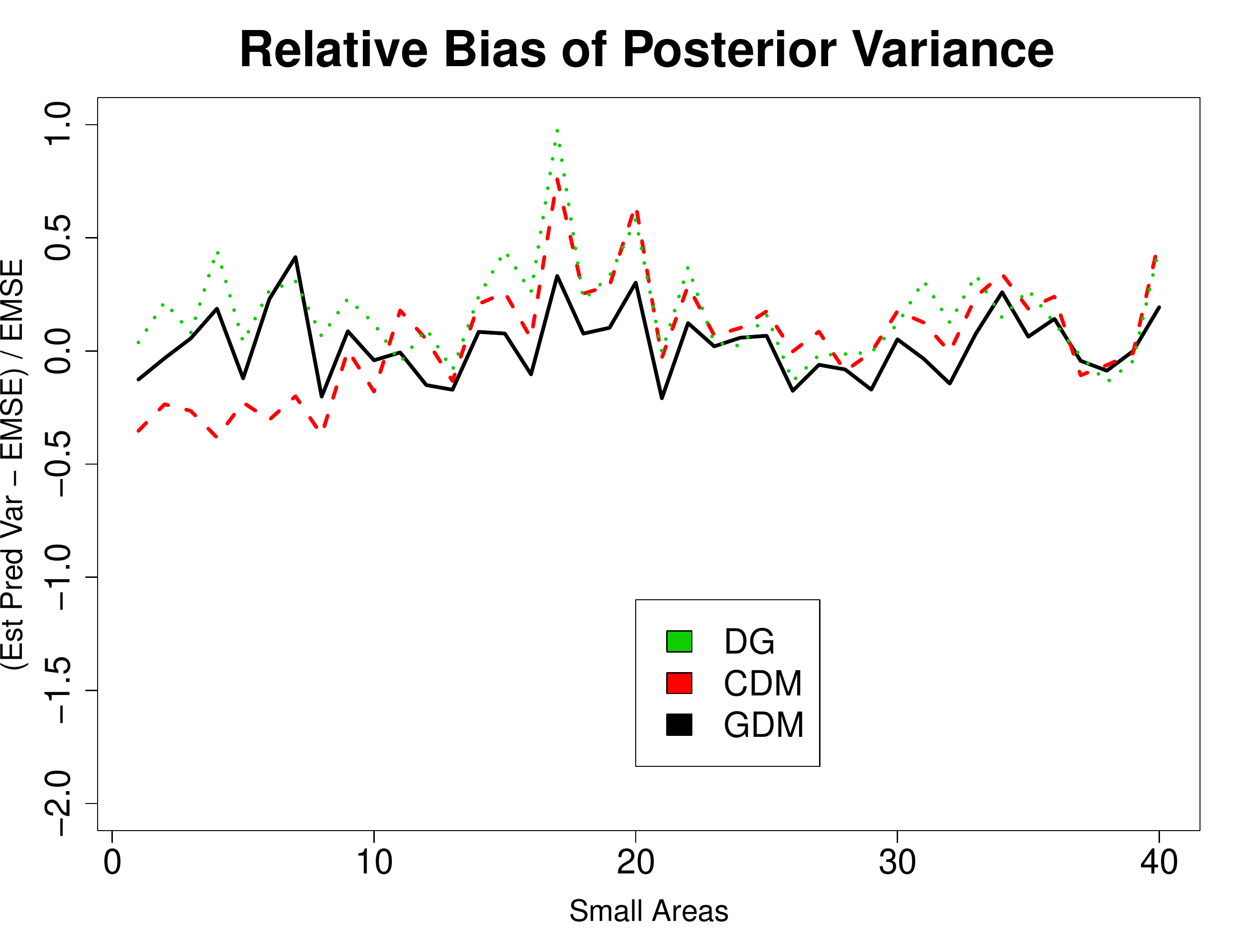}  \\
				\hspace{-.5in}\raisebox{-3cm}{3\% $N(5,5^2)$\hspace{.075cm}}     & \includegraphics[scale=.275,angle=270] {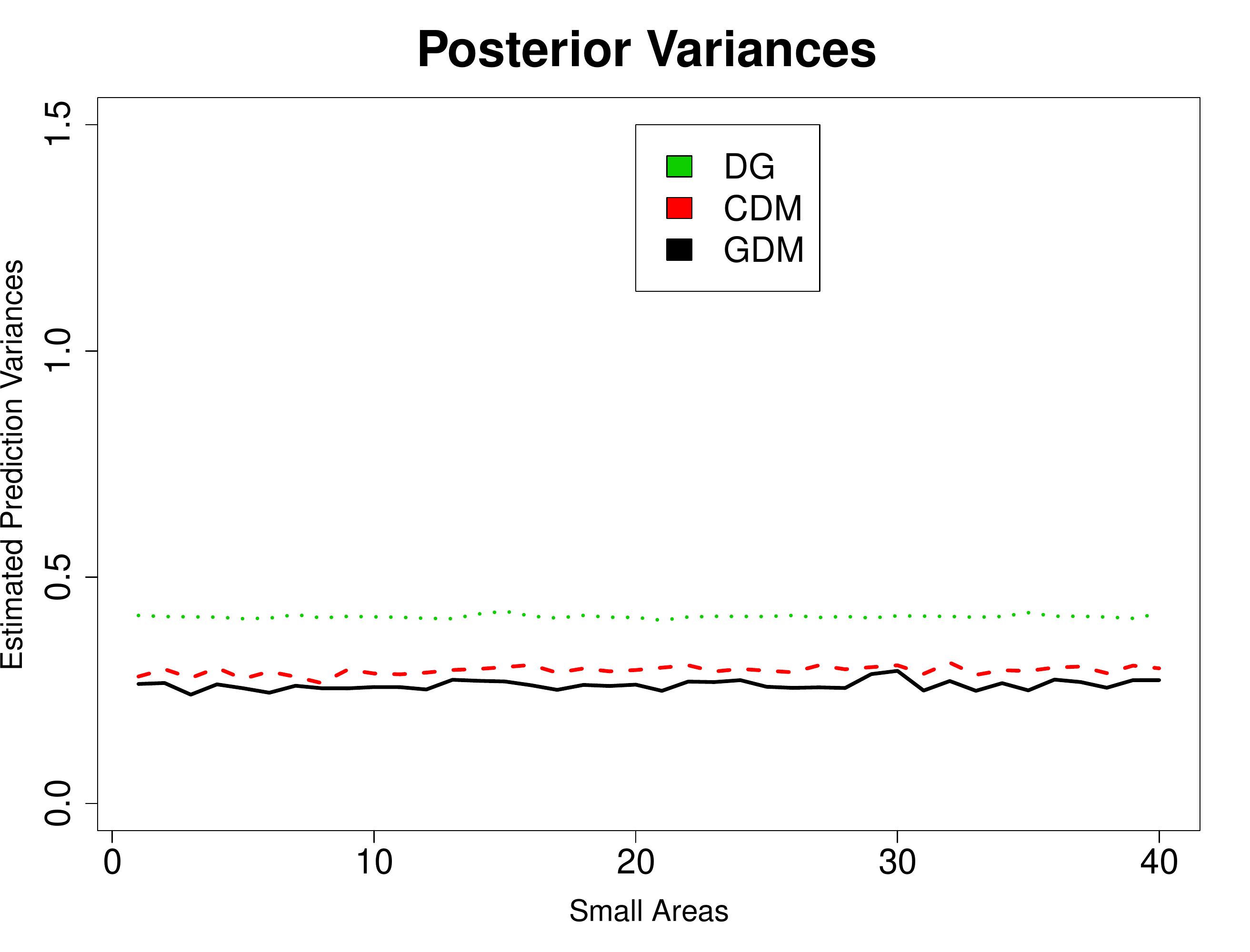} & \includegraphics[scale=.275,angle=270]{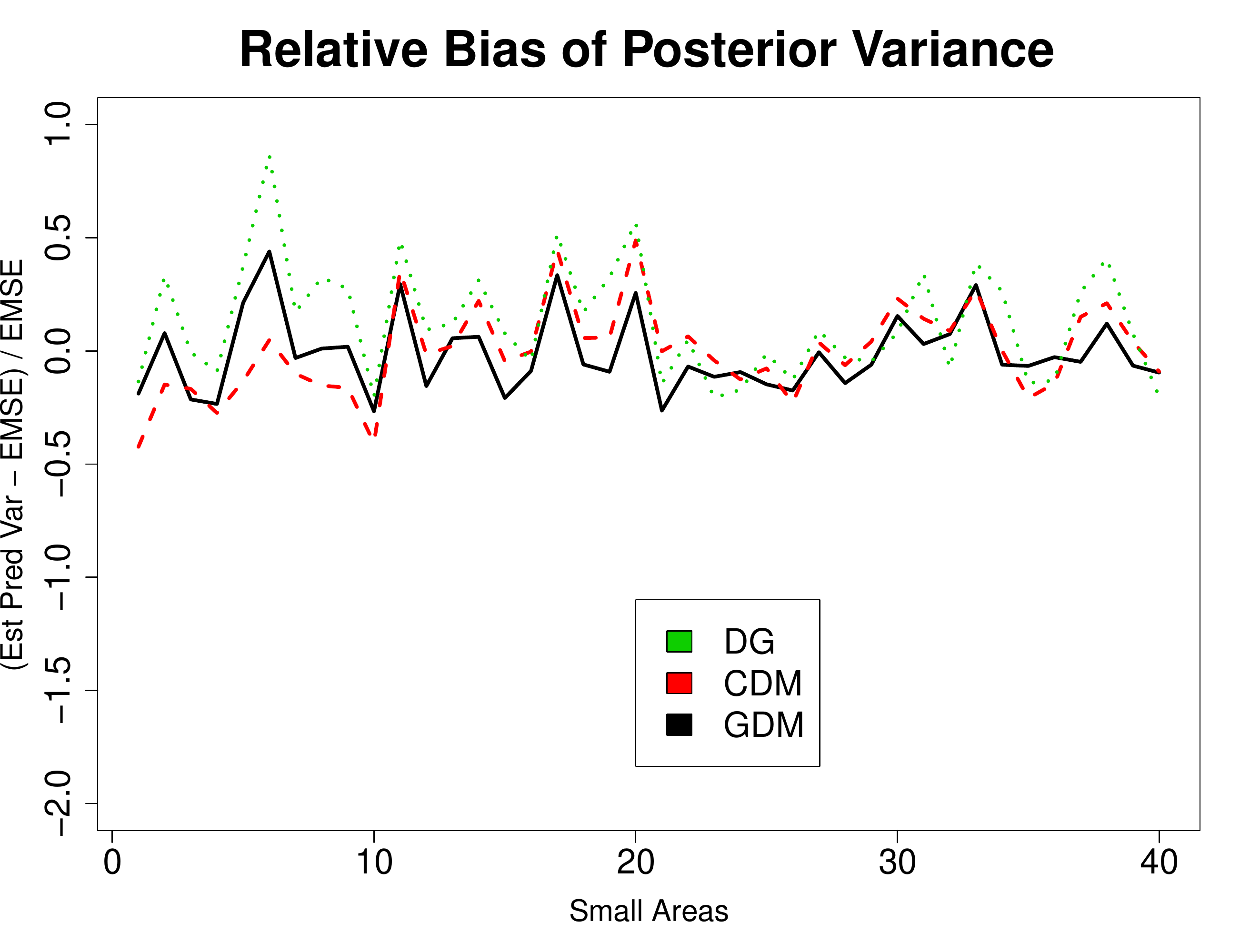}  \\
				\hspace{-.5in}\raisebox{-3cm}{$t_{(4)}$\hspace{.075cm}}     & \includegraphics[scale=.275,angle=270] {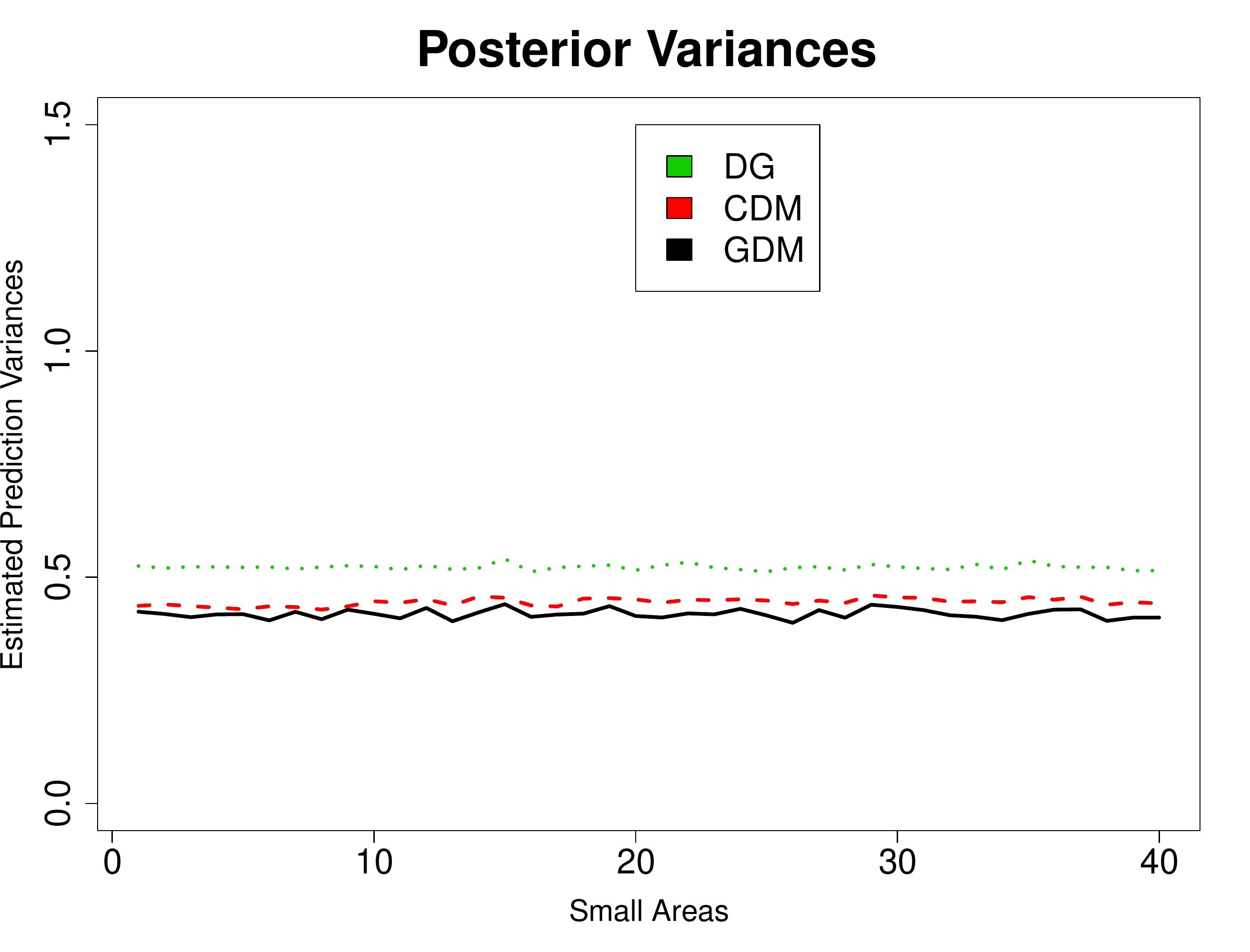} & \includegraphics[scale=.275,angle=270]{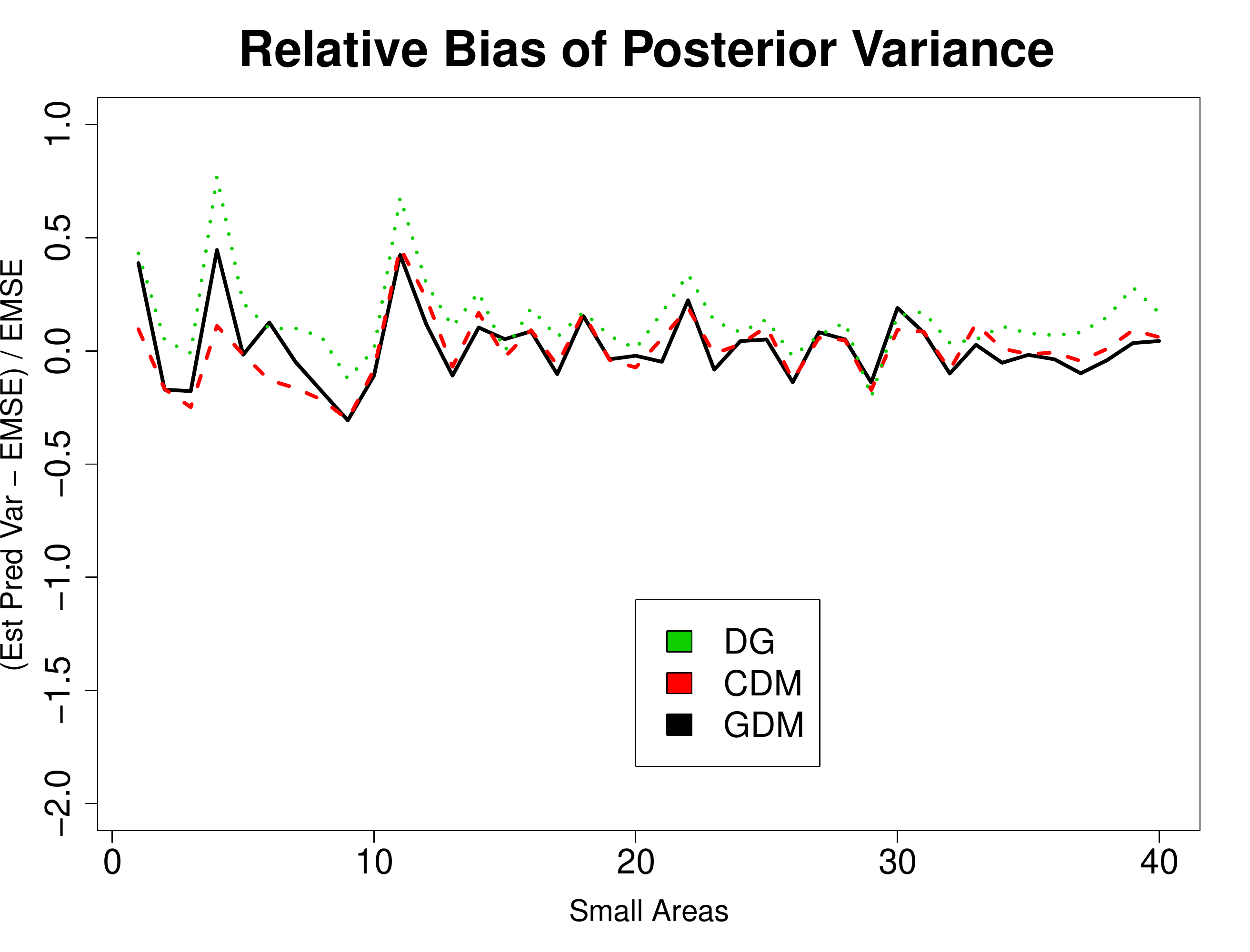}  \\
			\end{tabular}
			\caption{Plot of posterior variances and their empirical relative biases }\label{tout:EMSE4}
		\end{center}
	\end{figure}
	
	\clearpage
	
	\thispagestyle{empty}
	\begin{figure}[h]
		\vspace{-.5cm}
		\begin{center}
			\begin{tabular}{ccc}
				\hspace{-.5in}\raisebox{-3cm}{40\% $N(0,5^2)$\hspace{.075cm}}     & \includegraphics[scale=.275,angle=270] {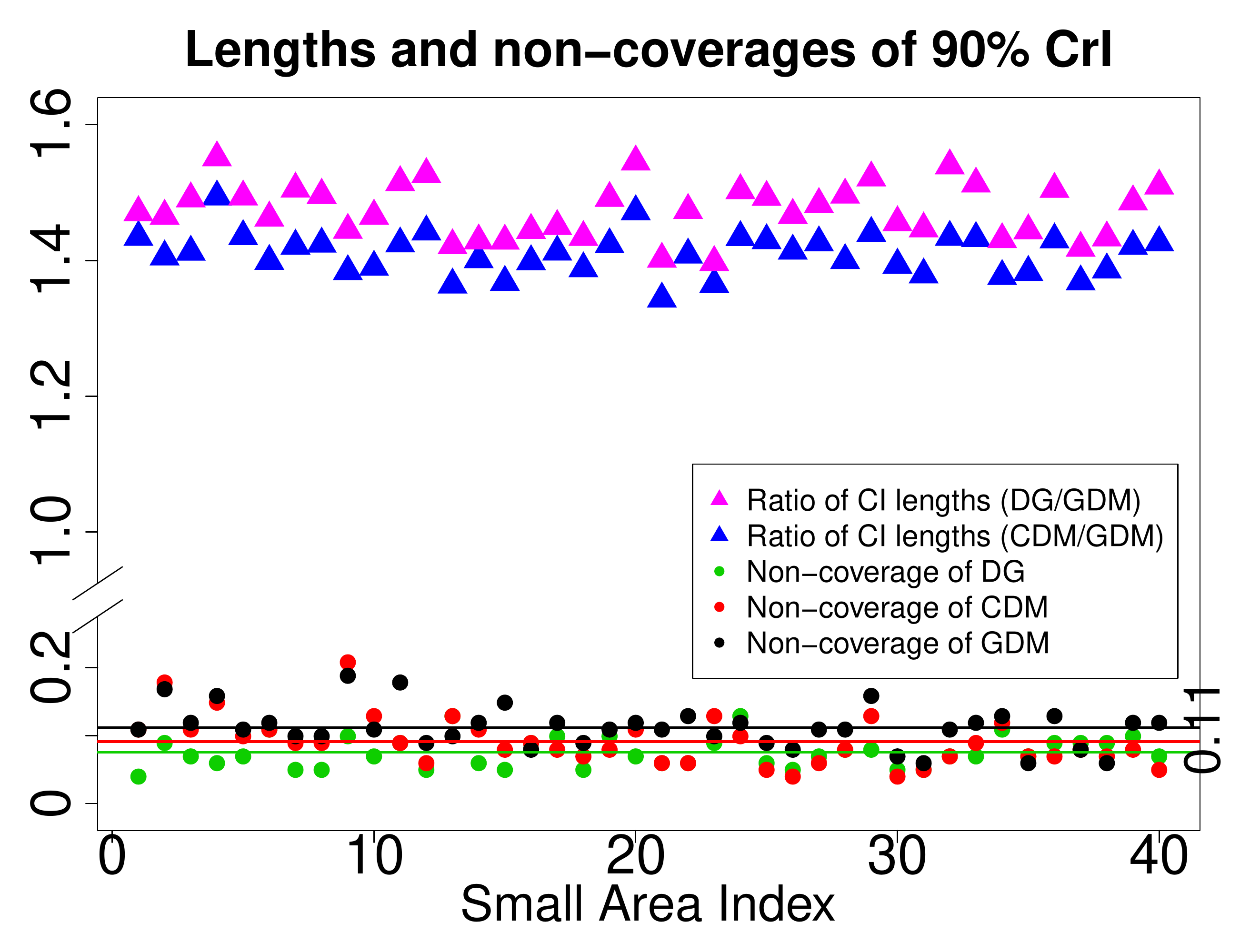} & \includegraphics[scale=.275,angle=270]{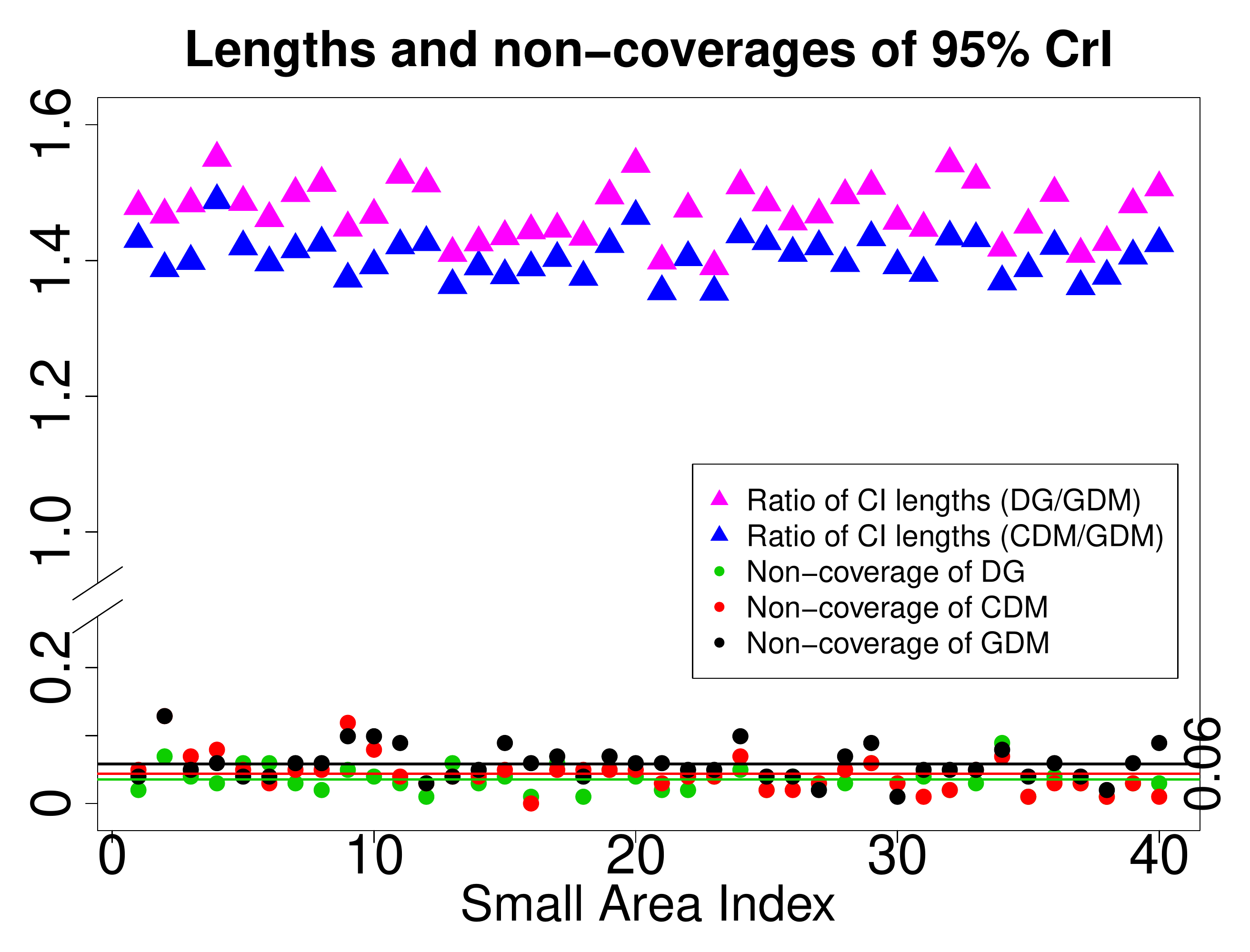}  \\
				\hspace{-.5in}\raisebox{-3cm}{10\% $N(0,5^2)$\hspace{.075cm}}     & \includegraphics[scale=.275,angle=270] {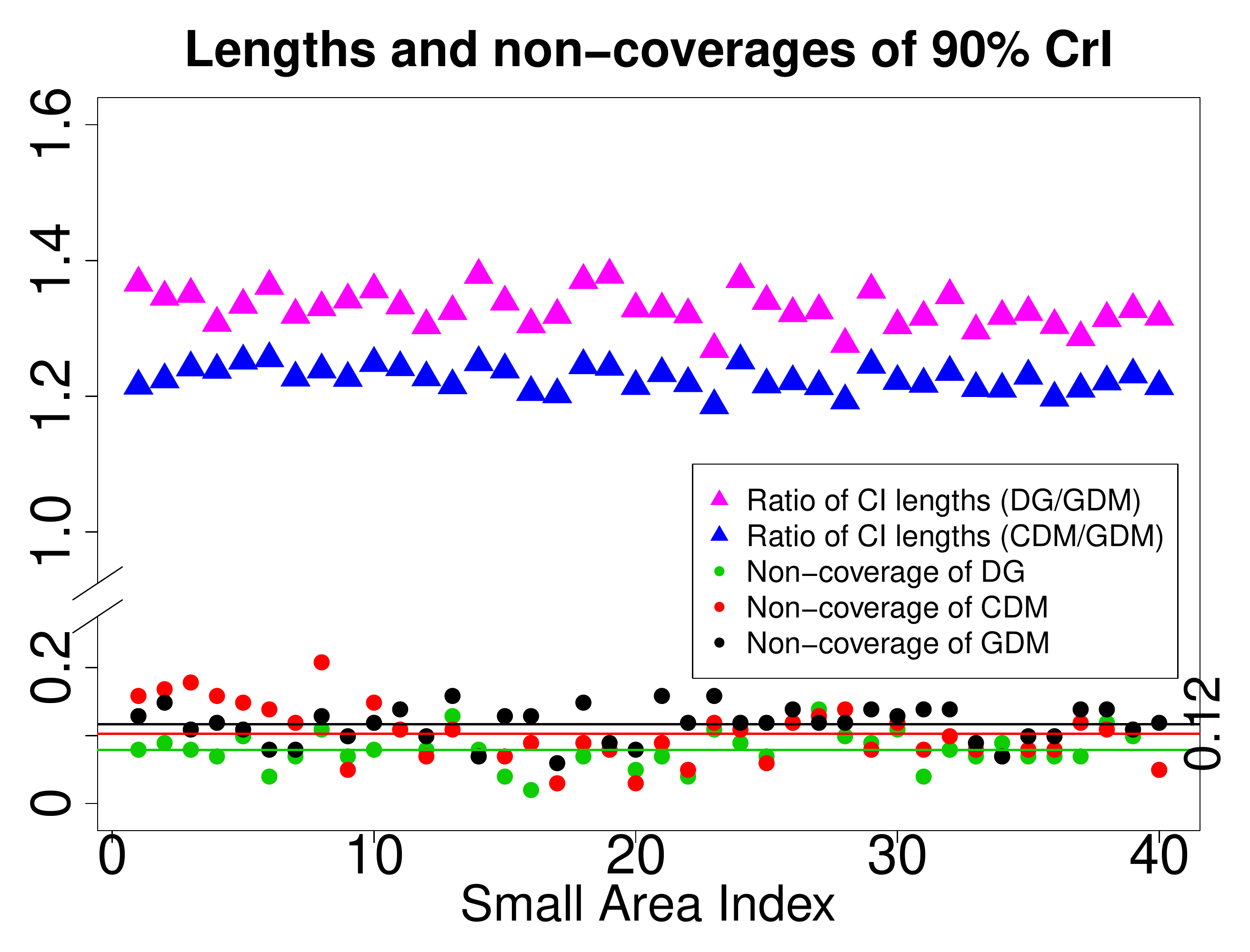} & \includegraphics[scale=.275,angle=270]{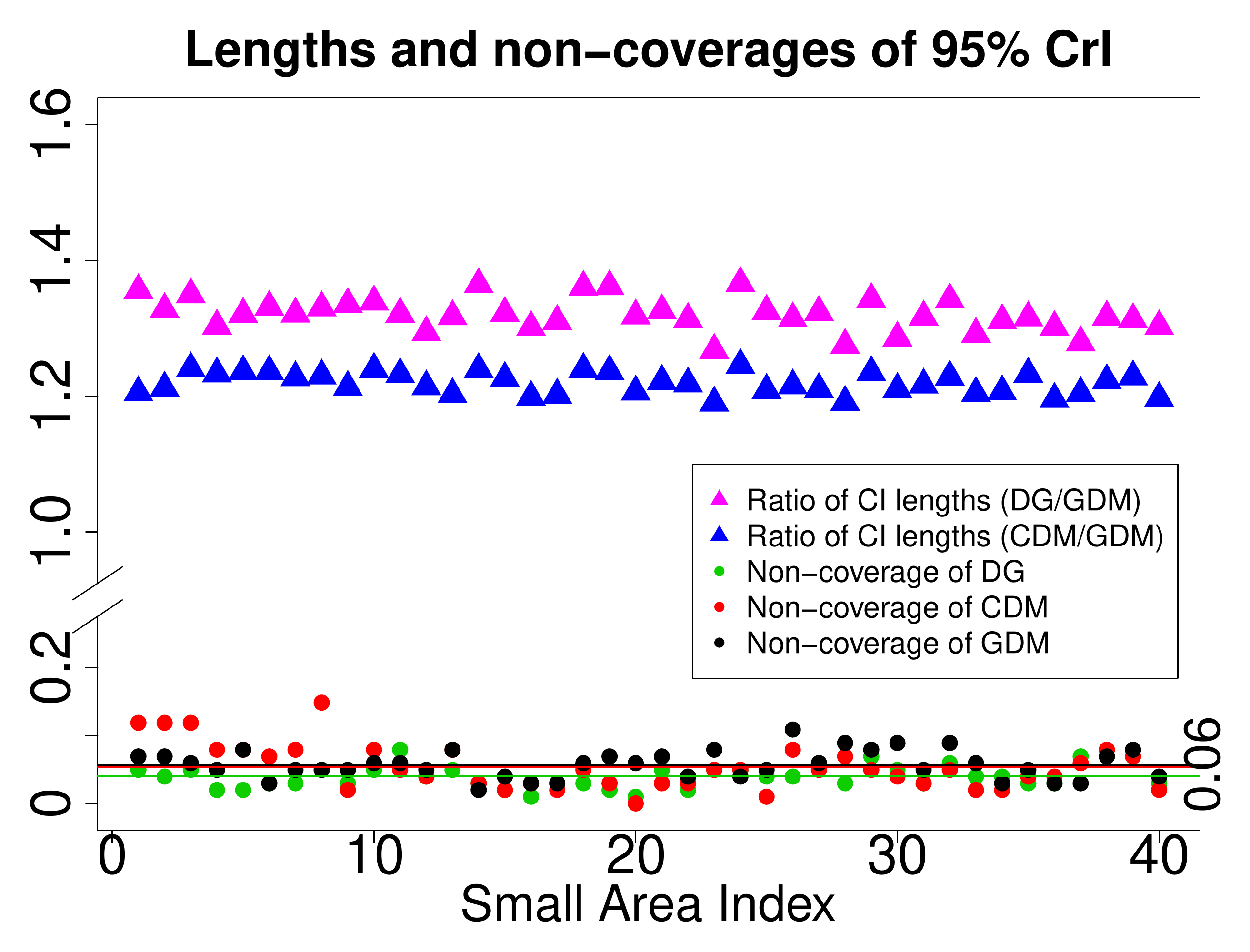}  \\
				\hspace{-.5in}\raisebox{-3cm}{3\% $N(5,5^2)$\hspace{.075cm}}     & \includegraphics[scale=.275,angle=270] {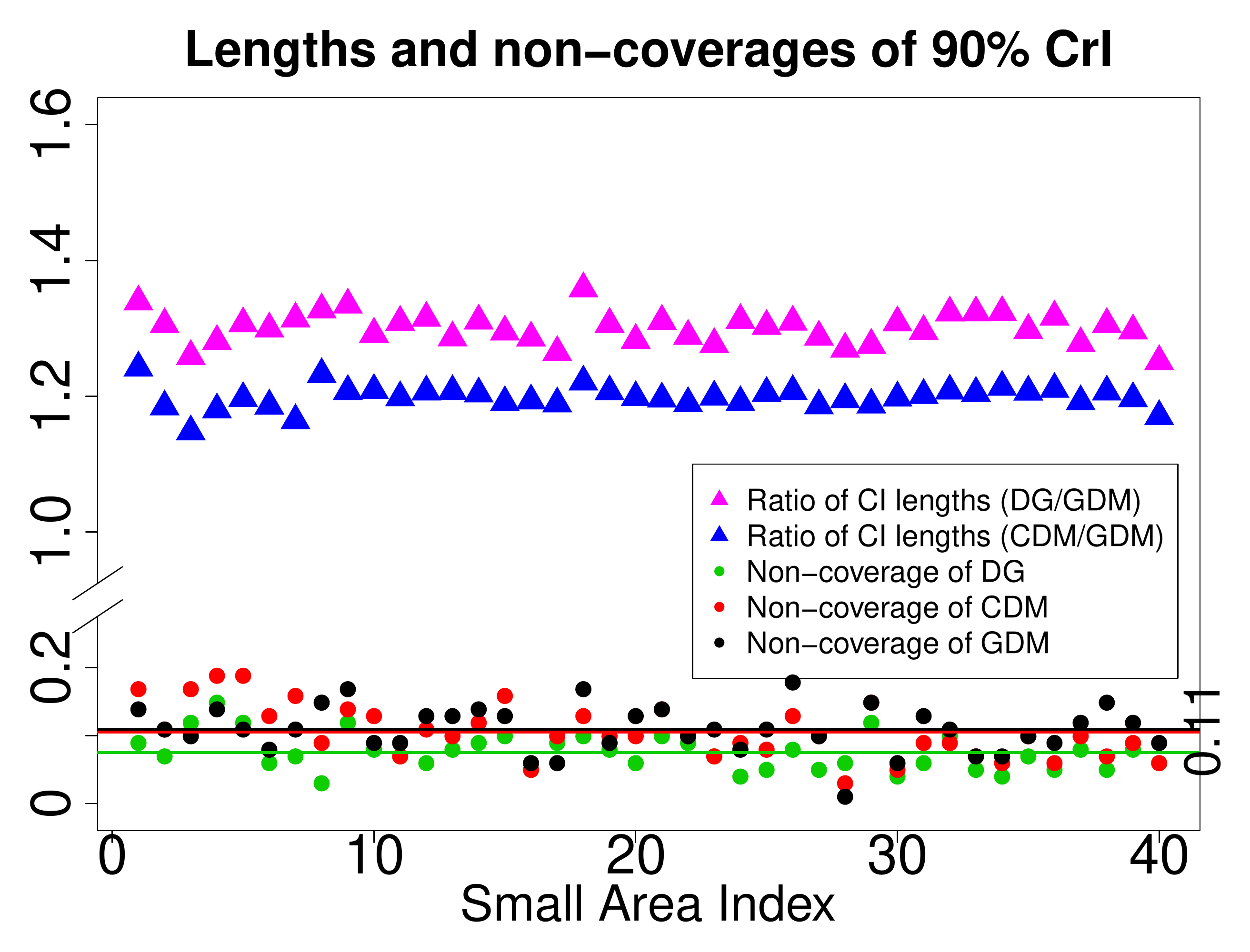} & \includegraphics[scale=.275,angle=270]{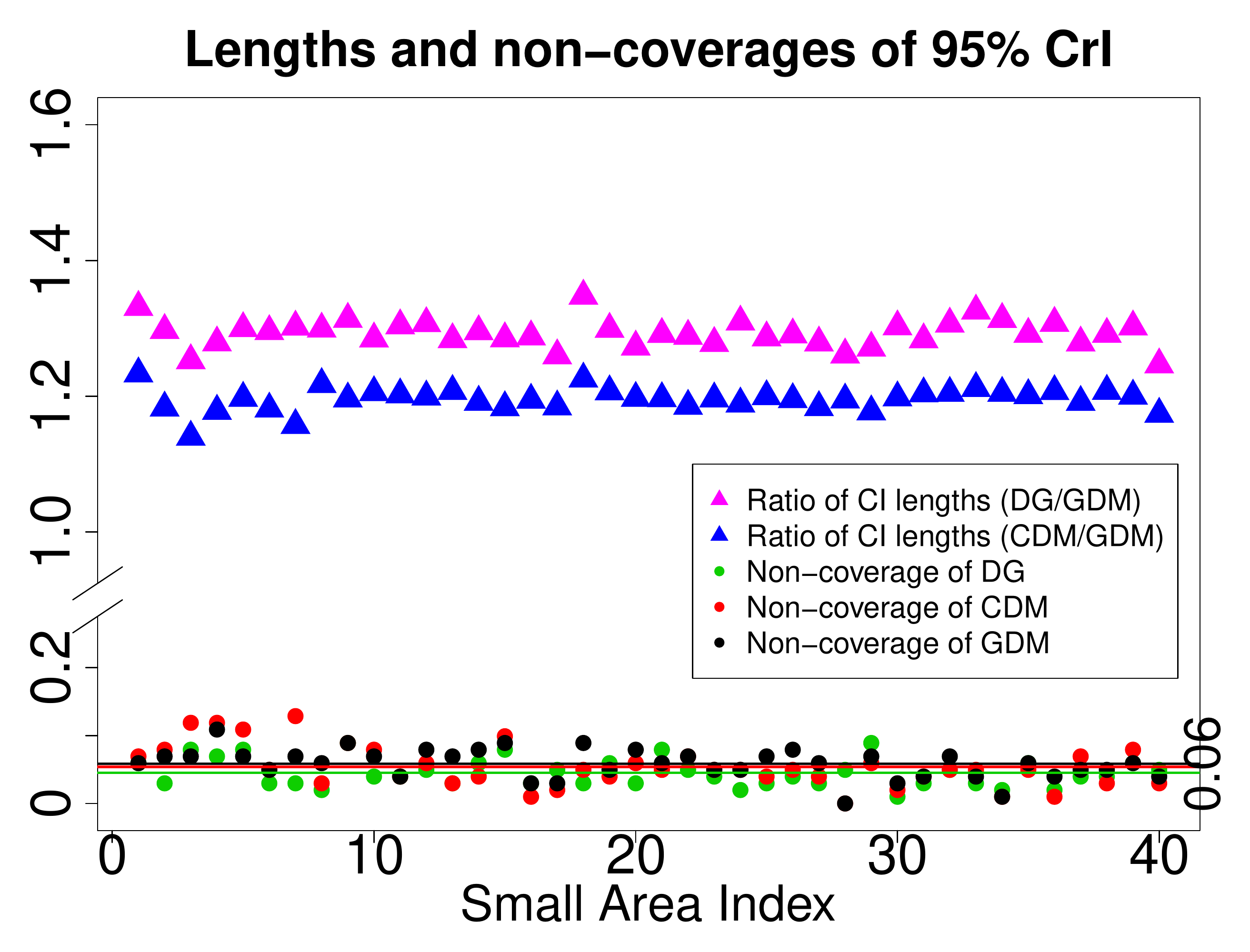}  \\
				\hspace{-.5in}\raisebox{-3cm}{$t_{(4)}$\hspace{.075cm}}     & \includegraphics[scale=.275,angle=270] {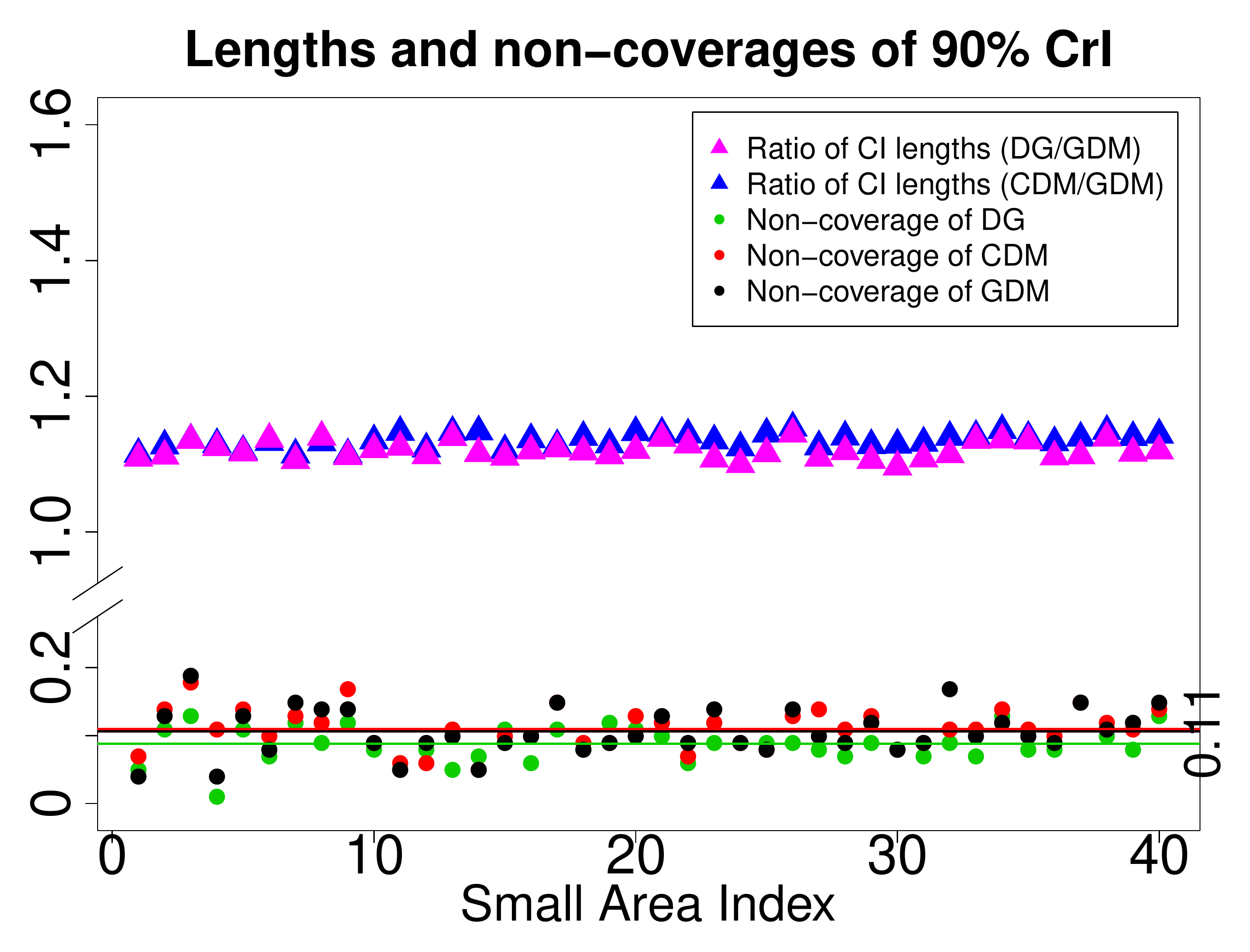} & \includegraphics[scale=.275,angle=270]{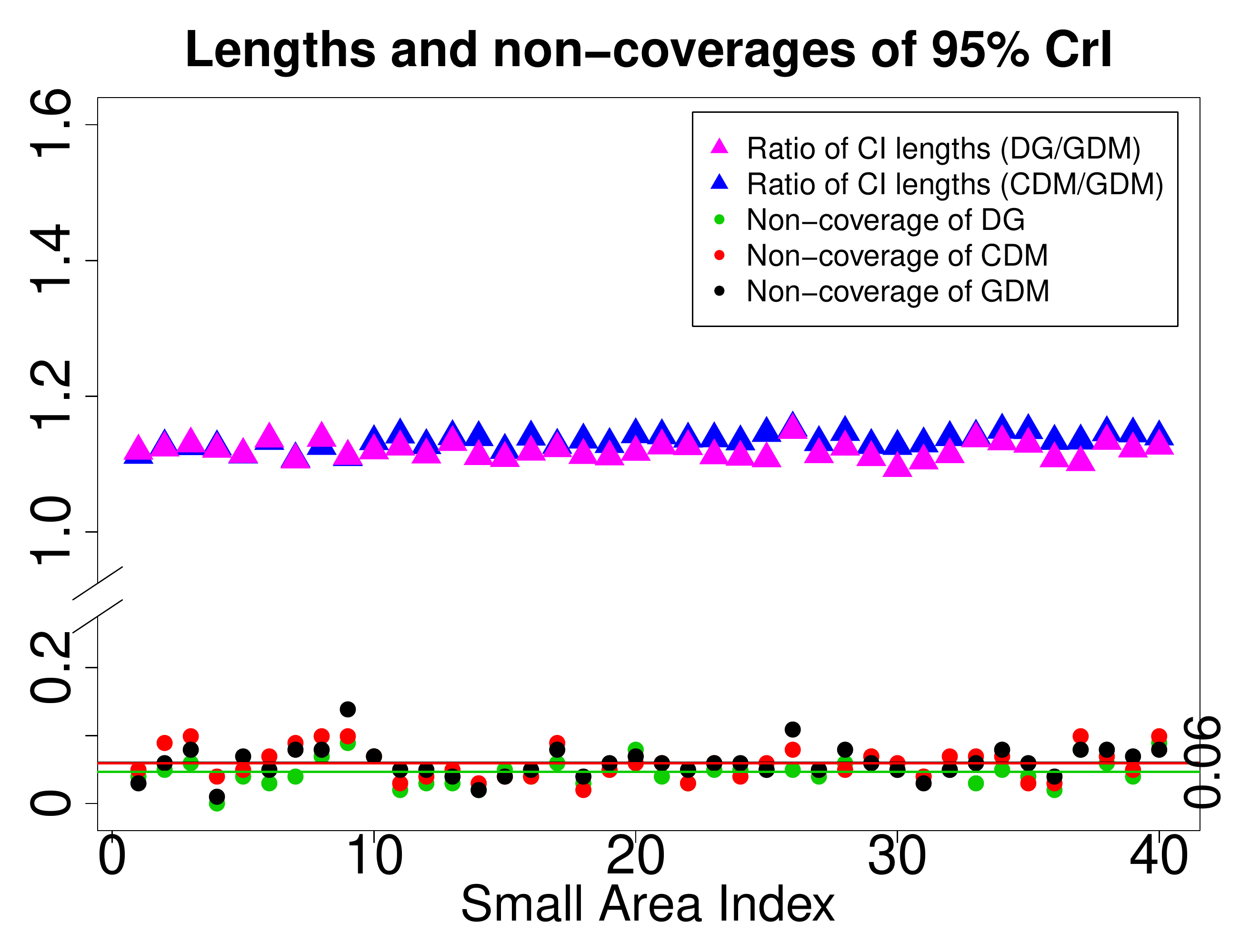}  \\
			\end{tabular}
			\caption{Plot of lengths and non-coverages of credible intervals (CrI)}\label{tout:CI2}
		\end{center}
	\end{figure}
	\clearpage

	\bigskip\noindent For each simulation setup, we simulate $S=100$ populations. For the $s^{th}$ simulated population, where $s=1,\ldots,S$, we compute the true small area means $\theta_i^{(s)}$. We denote the predictors of small area means calculated using HB methods as $\hat{\theta}^{(s)}_i$ and the variances of those predictors as $V_i^{(s)}$. For each HB method, given the predicted small area means $\hat\theta^{(s)}_i$, we calculate empirical biases as $eB_i=\frac{1}{S}\sum_{s=1}^{S}\big(\hat{\theta}_i^{(s)}-\theta_i^{(s)}\big)$ and empirical mean squared errors as $eM_i=\frac{1}{S}\sum_{s=1}^{S}\big(\hat{\theta}_i^{(s)}-\theta_i^{(s)}\big)^2$. In Figure 3, we provide plots of empirical biases and empirical mean squared errors (MSEs) for HB predictors considered. None of the HB predictors shows signs of systematic bias. However, in the simulation setup where $p_e=0.6$, the empirical biases of the GDM HB predictors seem to have smaller variability than the empirical biases of the other two HB predictors. In the case of 3\% contamination in $e_{ij}$ or where $e_{ij}$ is determined by a $t$-distribution, the three models perform equally well in producing small MSEs. In the case of 10\% contamination, the MSEs of the CDM and GDM HB predictors are approximately equal for most of the small areas but smaller than the DG model prediction. The most substantial difference among the three models results in the case where $p_e=0.6$. Here, the GDM predictor has the lowest MSEs overall of the three methods, followed by the CDM predictor and then by the DG predictor. Moreover, the GDM model MSEs stay generally stable across all small areas.

	\bigskip\noindent Next, Figure 4 shows posterior variances $V_i^{(s)}$ for 40 small areas and the relative biases of those variances, calculated as $RE_{V}=\{(1/S)\sum_{s=1}^SV_i^{(s)}-eM_i\}/eM_i$. The CDM and GDM predictors seem to enjoy lower posterior variances than the DG model. Furthermore, as the amount of contamination increases, the GDM model also produces lower posterior variances than the CDM model. The differences between the three models become more pronounced as contamination increases. The DG model also displays a mild tendency toward positive relative bias when calculating posterior variance. The CDM and GDM models do not show systematic bias in calculations of posterior variance and overall perform equally well.	

	\bigskip\noindent Figure 5 shows the empirical non-coverage probabilities of 90\% and 95\% credible intervals of small area means $\theta_i$. \bt{For each simulation setup, we also use solid horizontal lines to show the mean non-coverage probability of the credible intervals produced by each method.} For a Bayesian method, we compute our 90\% credible interval $I^{(s)}_{i,90}$ for $\theta_i$ by the 5th and 95th percentiles of the posterior distribution of $\theta_i$. We then calculate the non-coverage probabilty of this credible interval as $eC_{i,90}=\frac{1}{S}\sum_{s=1}^S I[\theta_i^{(s)}\not\in I^{(s)}_{i,90}]$. The same calculations are done for 95\% credible intervals using the 2.5th and 97.5th percentiles. \bt{The label on the right axis of each plot is the non-coverage probability of the GDM credible intervals. The plots also show two ratios which compare the lengths of the DG and CDM credible intervals to those of the GDM credible intervals.} We denote the length of the 90\% credible interval $I^{(s)}_{i,90}$ as $L^{(s)}_{i,90}$. The empirical average length of the credible interval of $\theta_i$ for a specific HB method is then computed as $\bar{L}_{i,90}=\frac{1}{S}\sum_{s=1}^S L_{i,90}^{(s)}$. Again, this calculation is repeated for the 95\% credible intervals. We see the credible intervals produced by the DG method consistently have the lowest non-coverage probabilities for each simulation, compared to the CDM and GDM intervals. {We also observe that the credible intervals by the DG HB model are on average larger than those developed from the other two models, except for the $t$-distributed $e_{ij}$ scenario where the DG and CDM credible intervals have similar lengths. Though the DG credible intervals most often capture the true value $\theta_i^{(s)}$ and have low non-coverage probabilities, they are longer than the GDM credible intervals, which closely attain the target coverage probability. While CDM and GDM intervals have similar non-coverage probabilities and nearly achieve the target when $e_{ij}$ is generated from a $t$-distribution, the ratios of average lengths (CDM/GDM) are consistently higher than one when greater levels of contamination are introduced into the population, indicating that the narrower GDM credible intervals are as successful as the CDM credible intervals in capturing the true values. }

	\bigskip\noindent At 3\% contamination of $e_{ij}$ from the secondary distribution ($p_e=0.97$), the non-coverage probabilities of the GDM and CDM credible intervals remain approximately equal, but the 90\% and 95\% intervals produced by the CDM model are up to 5\% greater in length than their respective GDM measures. When 10\% of $e_{ij}$ come from a secondary distribution ($p_e=0.90$), the non-coverage probabilities of the credible intervals found from the CDM approach are slightly lower than those found from the GDM model, but the CDM credible intervals are also about 10\% greater in length than their respective GDM measures. When $e_{ij}$ comes from a primary distribution with probability $p_e=0.60$, the CDM model credible intervals are about 40\% longer than those given by the GDM model but continue to have a slightly lower non-coverage probability. We note that the non-coverage probabilities of the GDM credible intervals seem to be consistent across all simulation setups. In contrast, the non-coverage probabilities of the CDM credible intervals appear to decrease when the concentration of $e_{ij}$ from a secondary distribution increases, but the CDM credible intervals become wider relative to the GDM credible intervals in higher contamination cases.

	\section{Conclusion}\label{sec:conclusion}
	Since Battese et al. (1988) introduced the NER model it has been the basis for many important developments in small area estimation for unit-level data. Datta and Ghosh (1991) applied HB methods to the NER model to develop Bayesian inference for small area means. This approach, however, is not robust in the presence of outliers or under non-normality of unit-level errors. The HB method proposed by Chakraborty et al. (2018), which also relied on an HB approach to the NER model, built upon the work of Datta and Ghosh (1991) to accommodate populations contaminated with outliers due to unit-level errors.
	
	\bigskip\noindent The CDM model is robust in the presence of outliers, but not under circumstances where the proportion of unit-level errors from the secondary distribution is fairly large. In this paper, we propose an alternate HB approach to extend the NER model for more general cases where unit-level errors come from a mixture of two different normal distributions. Based on simulation studies, we find that the proposed model provides HB estimates with lower empirical MSEs, posterior variances and narrower credible intervals than the DG and CDM HB models. The consistent superior performance of the proposed model to the DG and CDM HB models regardless of the presence of mixture in the unit-level error indicates that there is no loss to applying it to all data sets.

	\section*{Acknowledgment}\label{sec:Ackon}
	The authors are thankful to Dr. Ray Chambers for providing the dataset used in Section~4.2.


	\clearpage
	\appendix
	\section{Appendix}
	\subsection{Integrability of joint posterior probability density function}\label{app:properdist}
	Chakraborty et al. (2018) show that the joint posterior density function of $\bm{\beta}, \sigma^2_1, \sigma^2_2, p_e,$ and $\sigma_v^2$ is proper. In particular, they show that the function
	\begin{eqnarray}
	L(\bm{\beta}, \sigma^2_1, \sigma^2_2, p_e,\sigma_v^2)\dfrac{I_{(\sigma_1^2<\sigma_2^2)}}{(\sigma_2^2)^2}
	\label{eq:CDMprop}
	\end{eqnarray}
	is integrable with respect to $\bm{\beta}, \sigma^2_1, \sigma^2_2, p_e,$ and $\sigma_v^2$, where $L(\bm{\beta}, \sigma^2_1, \sigma^2_2, p_e,\sigma_v^2)$ is the likelihood function based on the distribution $y_{ij},j=1,\ldots,n_1, i=1,\ldots,m$ obtained as the marginal distribution from (I)$-$(III) in Section 2.
	\bigskip\noindent Similar arguments show that  $L(\bm{\beta}, \sigma^2_1, \sigma^2_2, p_e,\sigma_v^2)\dfrac{I_{(\sigma_1^2\geq\sigma_2^2)}}{(\sigma_1^2)^2}$ is also integrable with respect to the same variables. Note that
	\begin{eqnarray*}
		\dfrac{I_{2^{-1}<p_e<1}}{(\sigma_1^2+\sigma_2^2)^2} & \leq \dfrac{1}{(\sigma_1^2+\sigma_2^2)^2} =  \dfrac{I_{(\sigma_1^2<\sigma_2^2)}+I_{(\sigma_1^2\geq\sigma_2^2)}}{(\sigma_1^2+\sigma_2^2)^2} \\[10pt]
		&=\dfrac{1}{(\sigma_2^2)^2}\bigg(\dfrac{\sigma_2^2}{\sigma_1^2+\sigma_2^2}\bigg)^2I_{(\sigma_1^2<\sigma_2^2)}+\dfrac{1}{(\sigma_1^2)^2}\bigg(\dfrac{\sigma_1^2}{\sigma_1^2+\sigma_2^2}\bigg)^2I_{(\sigma_1^2\geq\sigma_2^2)}
		&< \dfrac{I_{(\sigma_1^2<\sigma_2^2)}}{(\sigma_2^2)^2}+\dfrac{I_{(\sigma_1^2\geq\sigma_2^2)}}{(\sigma_1^2)^2}
	\end{eqnarray*}
	
	This implies,
	\begin{eqnarray}\label{eqn:appendix}
	L(\bm{\beta}, \sigma^2_1, \sigma^2_2, p_e,\sigma_v^2)\dfrac{I_{2^{-1}<p_e<1}}{(\sigma_1^2+\sigma_2^2)^2}
	&< L(\bm{\beta}, \sigma^2_1, \sigma^2_2, p_e,\sigma_v^2)\bigg(\dfrac{I_{(\sigma_1^2<\sigma_2^2)}}{(\sigma_2^2)^2}+\dfrac{I_{(\sigma_1^2\geq\sigma_2^2)}}{(\sigma_1^2)^2}\bigg)
	\end{eqnarray}
	
	Since LHS of (4) is bounded above by two integrable functions, it is also integrable.

	\subsection{Simulation results with no contamination of unit-level errors}\label{app:nocontam}
	\begin{figure}[h]
		\begin{center}
			\begin{tabular}{cc}
				\includegraphics[scale=.275,angle=270] {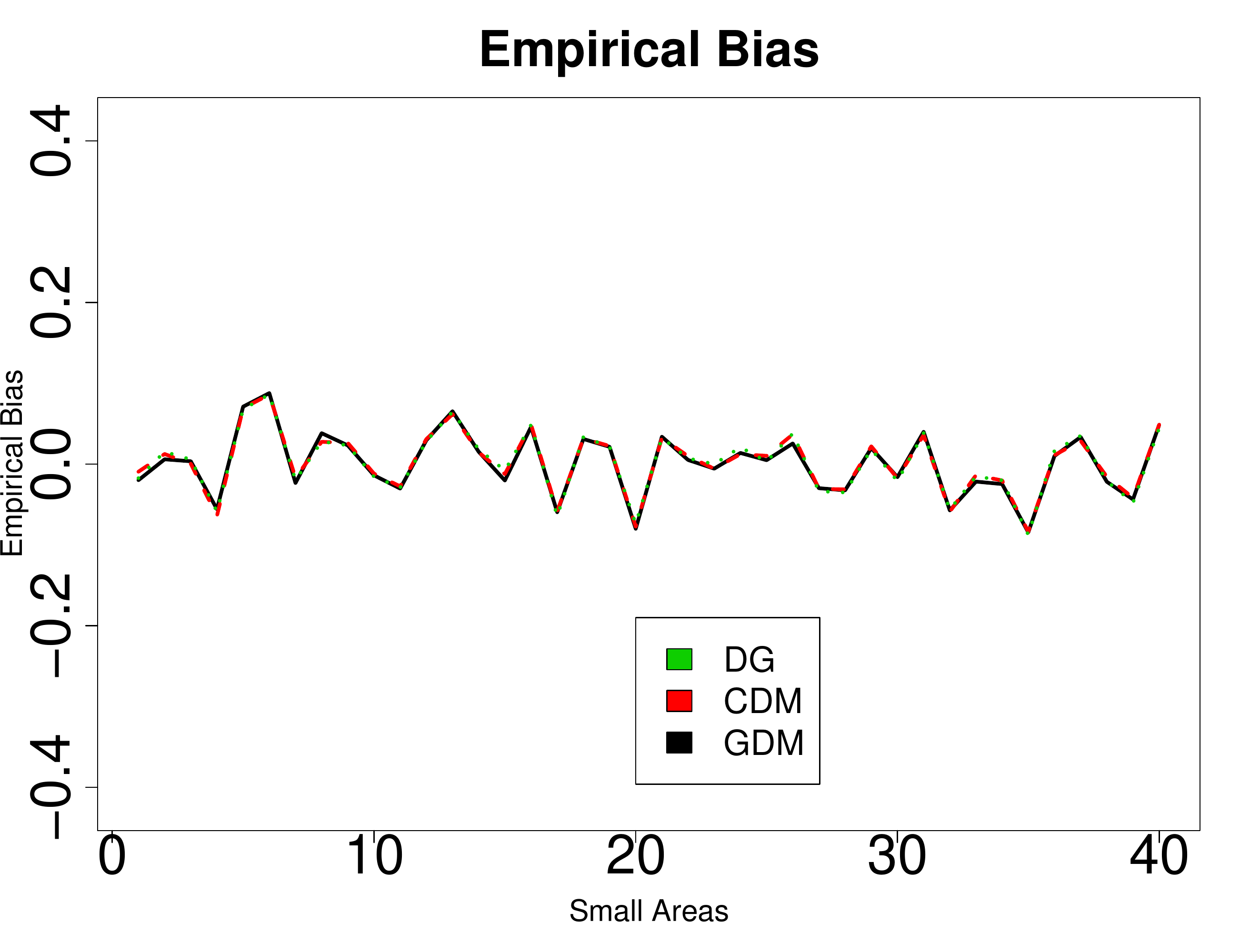} & \includegraphics[scale=.275,angle=270]{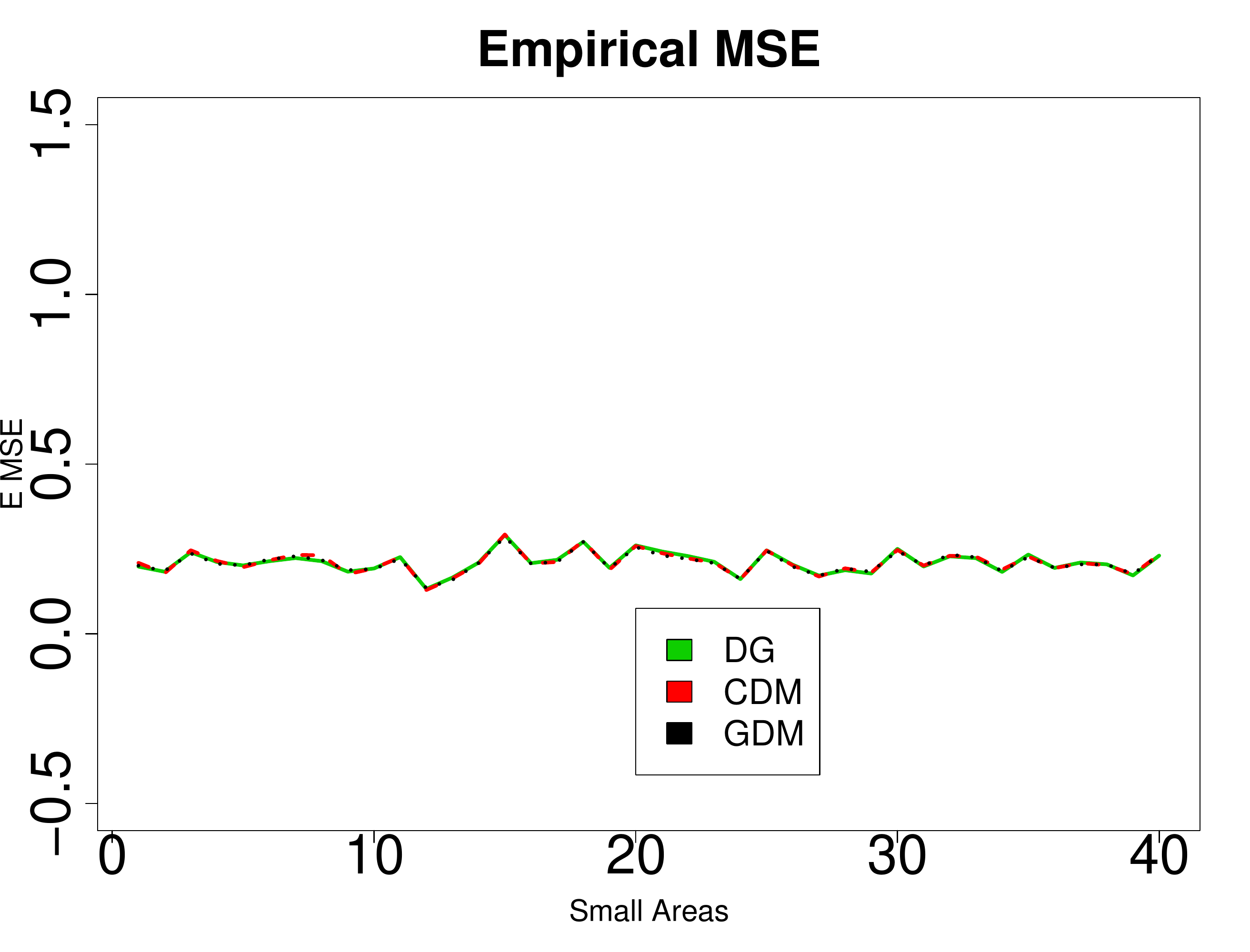}  \\
				\includegraphics[scale=.275,angle=270] {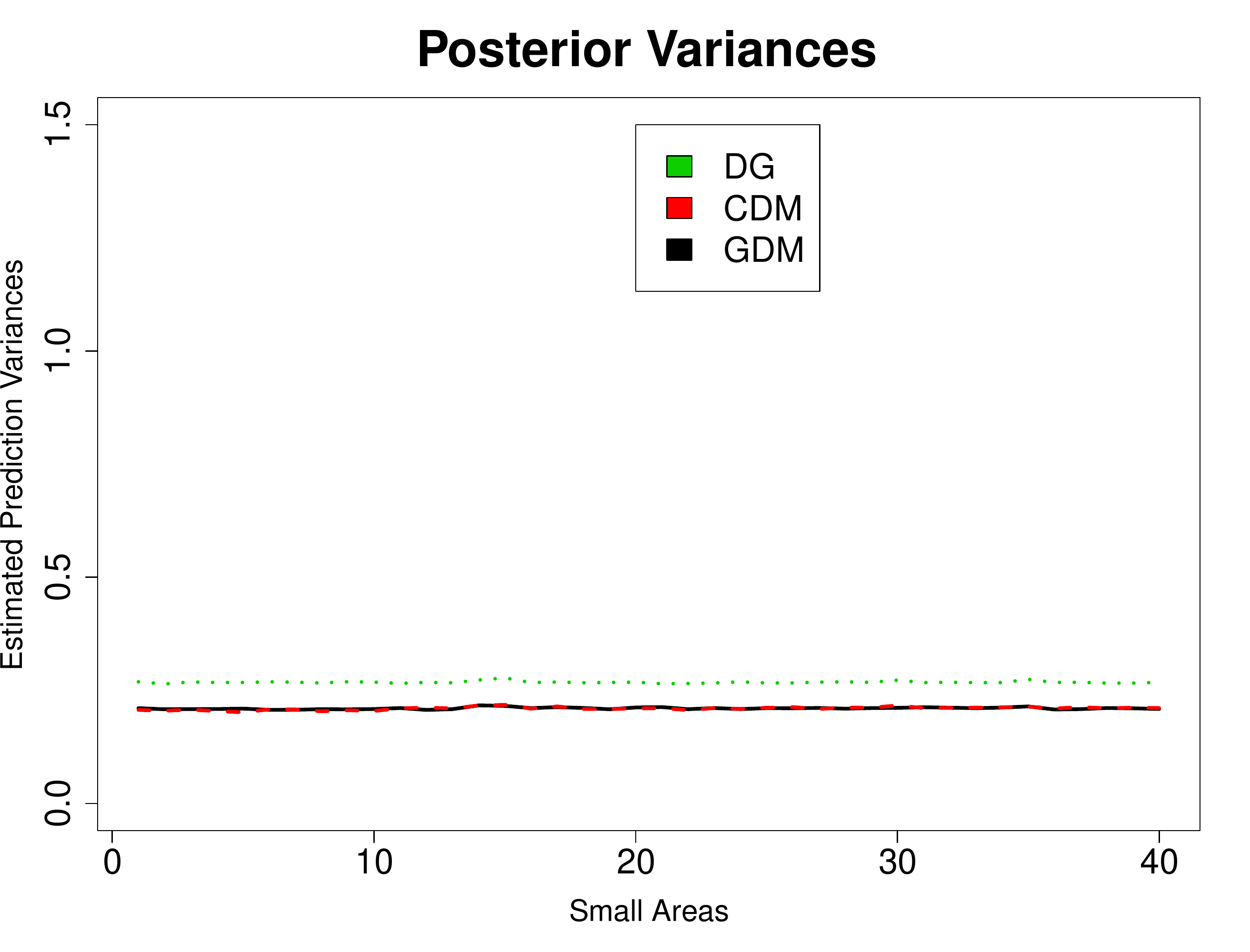} & \includegraphics[scale=.275,angle=270]{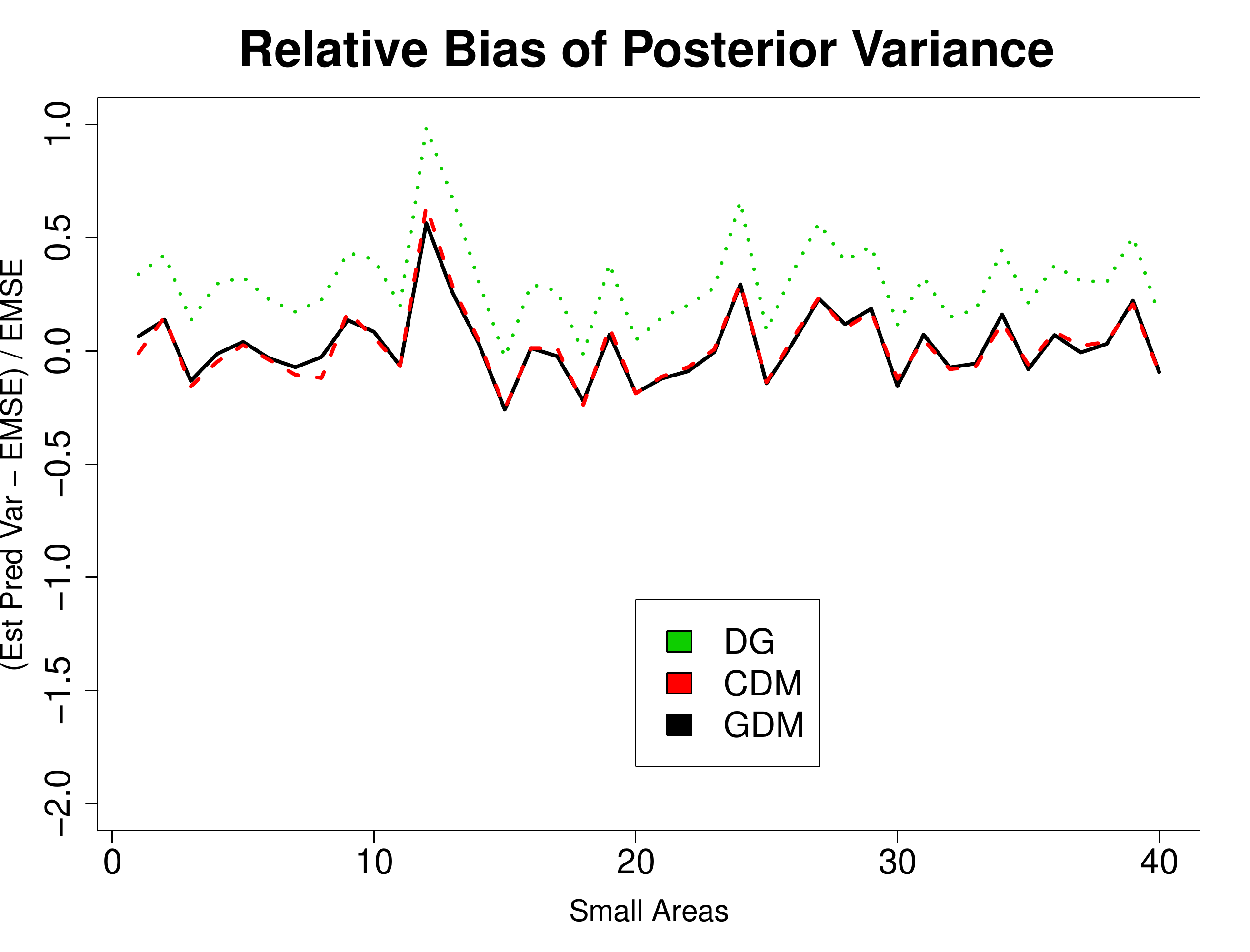}  \\
				\includegraphics[scale=.275,angle=270] {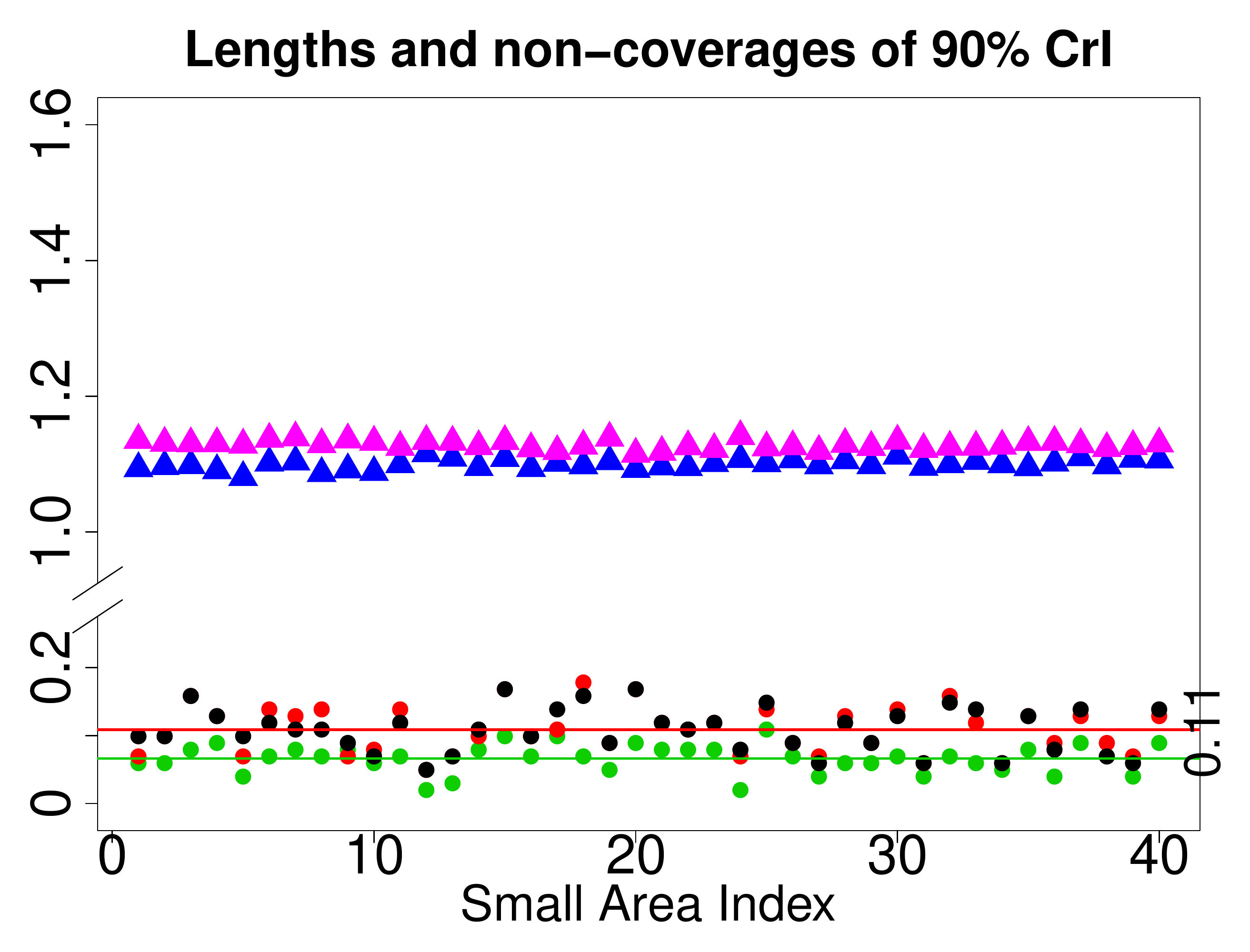} & \includegraphics[scale=.275,angle=270]{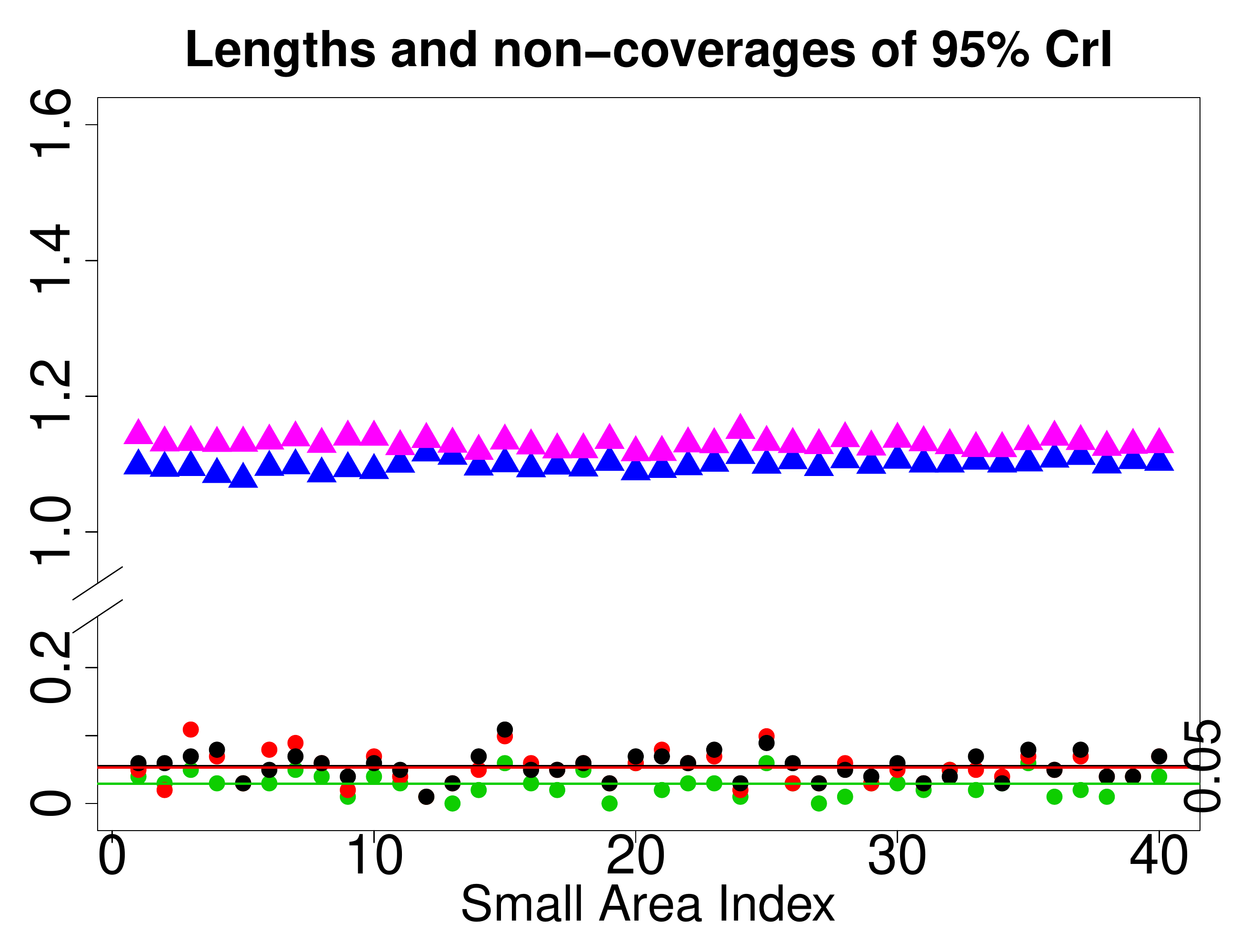}  \\
			\end{tabular}
			\caption{Plots of various measures of $\hat{\theta}$s when no unit-level contamination is present}\label{tout:nocontam}
		\end{center}
	\end{figure}
	\clearpage

	\clearpage
	\section*{Supplementary Materials}
	\subsection*{A priori probabilities}\label{app:apriori}
	
	The a priori probability that a unit is from the secondary population is
	\begin{eqnarray*}
		P(z_{ij}=0) = \int P(z_{ij}=0|p_e) \pi (p_e) dp_e  = \int_{\frac{1}{2}}^1 (1-p_e) 2 dp_e = -(1-p_e)\Big|_{\frac{1}{2}}^1=\frac{1}{4}
	\end{eqnarray*}
	and that a unit is from the primary population is
	\begin{eqnarray*}
		P(z_{ij}=1) &=& 1-P(z_{ij}=0) =\frac{3}{4}.
	\end{eqnarray*}

	\subsection*{Joint posterior distribution}\label{app:jointdens}
	Implementation of the model through Gibbs sampling requires the conditional distributions to be derived from the full joint density. The joint density of $\bm{y}=\{y_{ij};j=1,...,n_i, i=1,...,m\}, \bm{z}=\{z_{ij};j=1,...,n_i, i=1,...,m\},$ and $\bm{v}=\{v_1,...v_m\}$ is written as:
	{\small
		\begin{eqnarray*}
			f(\bm{y},\bm{v},\bm{\beta},\bm{z},\sigma_1^2,\sigma_2^2,\sigma_v^2,p_e) & \propto \\
			&\hspace{-1in} \dfrac{\exp\bigg[ -\dfrac{1}{2}\displaystyle\sum_{i=1}^{m}\sum_{j=1}^{n_i}\bigg(\dfrac{(y_{ij}-\bm{x_{ij}^T}\bm{\beta}-v_i)^2}{\sigma_1^2}z_{ij}+
				\dfrac{(y_{ij}-\bm{x_{ij}^T}\bm{\beta}-v_i)^2}{\sigma_2^2}(1-z_{ij})\bigg)\bigg]}{(\sigma_1^2)^{\dfrac{1}{2}\displaystyle\sum_{i=1}^{m}\sum_{j=1}^{n_i}z_{ij}}(\sigma_2^2)^{\dfrac{1}{2}\displaystyle\sum_{i=1}^{m}\sum_{j=1}^{n_i}(1-z_{ij})}} \\[10pt]
			&\hspace{-1in} \times \dfrac{\exp\bigg(-\dfrac{1}{2}\displaystyle\sum_{i=1}^{m}\dfrac{v_i^2}{\sigma_v^2}\bigg)}{(\sigma_v^2)^{m/2}}
			\times p_e^{\displaystyle\sum_{i=1}^{m}\displaystyle\sum_{j=1}^{n_i}z_{ij}}(1-p_e)^{\displaystyle\sum_{i=1}^{m}\displaystyle\sum_{j=1}^{n_i}(1-z_{ij})} \times I_{\{\frac{1}{2}<p_e<1\}}
			\times \dfrac{1}{{(\sigma_1^2+\sigma_2^2)}^2}
		\end{eqnarray*}
	}
	
	\bigskip\noindent To facilitate Gibbs sampling, we re-parametrize $\sigma_2^2=\eta\sigma_1^2$. The joint posterior distribution of the reparametrized model is:
	\begin{eqnarray*}
		f(\bm{y},\bm{v},\bm{\beta},\bm{z},\sigma_1^2,\eta,\sigma_v^2,p_e) & \propto \dfrac{\exp\bigg[ -\dfrac{1}{2}\displaystyle\sum_{i=1}^{m}\displaystyle\sum_{j=1}^{n_i}\bigg(\dfrac{(y_{ij}-\bm{x_{ij}^T}\bm{\beta}-v_i)^2}{\sigma_1^2}z_{ij}+
			\dfrac{(y_{ij}-\bm{x_{ij}^T}\bm{\beta}-v_i)^2}{\eta\sigma_1^2}(1-z_{ij})\bigg)\bigg]}{(\sigma_1^2)^{\dfrac{n}{2}+1}(\eta)^{\dfrac{1}{2}\displaystyle\sum_{i=1}^{m}\displaystyle\sum_{j=1}^{n_i}(1-z_{ij})}} \\[10pt]
		&\hspace{-1in} \times \dfrac{\exp\bigg(-\dfrac{1}{2}\displaystyle\sum_{i=1}^{m}\dfrac{v_i^2}{\sigma_v^2}\bigg)}{(\sigma_v^2)^{m/2}}
		\times p_e^{\displaystyle\sum_{i=1}^{m}\displaystyle\sum_{j=1}^{n_i}z_{ij}}(1-p_e)^{\displaystyle\sum_{i=1}^{m}\displaystyle\sum_{j=1}^{n_i}(1-z_{ij})}
		\times \dfrac{I_{\{2^{-1}<p_e<1\}}}{(1+\eta)^2}
	\end{eqnarray*}
	
	
	\subsection*{Conditional distributions for Gibbs sampling}\label{app:gibbsdist}
	We calculate the conditional distribution of each parameter using the joint pdf:
	\begin{eqnarray*}
		f(\bm{y},\bm{v},\bm{\beta},\bm{z},\sigma_1^2,\eta,\sigma_v^2,p_e) & \propto \dfrac{\exp\bigg[ -\dfrac{1}{2}\displaystyle\sum_{i=1}^{m}\displaystyle\sum_{j=1}^{n_i}\bigg(\dfrac{(y_{ij}-\bm{x_{ij}^T}\bm{\beta}-v_i)^2}{\sigma_1^2}z_{ij}+
			\dfrac{(y_{ij}-\bm{x_{ij}^T}\bm{\beta}-v_i)^2}{\eta\sigma_1^2}(1-z_{ij})\bigg)\bigg]}{(\sigma_1^2)^{\dfrac{n}{2}+1}(\eta)^{\dfrac{1}{2}\displaystyle\sum_{i=1}^{m}\displaystyle\sum_{j=1}^{n_i}(1-z_{ij})}} \\[10pt]
		&\hspace{-1in} \times \dfrac{\exp\bigg(-\dfrac{1}{2}\displaystyle\sum_{i=1}^{m}\dfrac{v_i^2}{\sigma_v^2}\bigg)}{(\sigma_v^2)^{m/2}}
		\times p_e^{\displaystyle\sum_{i=1}^{m}\displaystyle\sum_{j=1}^{n_i}z_{ij}}(1-p_e)^{\displaystyle\sum_{i=1}^{m}\displaystyle\sum_{j=1}^{n_i}(1-z_{ij})}
		\times \dfrac{I_{\{2^{-1}<p_e<1\}}}{(1+\eta)^2}
	\end{eqnarray*}
	\renewcommand{\theenumi}{(\Roman{enumi})}
	\begin{enumerate}
		\item
		$\bm{\beta}|\bm{y},\bm{v},\bm{z},\bm{\beta},\sigma_1^2,\eta,\sigma_v^2,p_e\sim
		N_p\bigg(S_\beta\displaystyle\sum_{i=1}^m\sum_{j=1}^{n_i}(y_{ij}-v_i)\bigg(\dfrac{z_{ij}}{\sigma_1^2}+\dfrac{1-z_{ij}}{\eta\sigma_1^2}\bigg)x_{ij},S_\beta\bigg)$
		\begin{eqnarray*}\text{where }S_\beta&=\bigg[\displaystyle\sum_{i=1}^{m}\sum_{j=1}^{n_i}x_{ij}x_{ij}^T\bigg(\dfrac{z_{ij}}{\sigma_1^2}+\dfrac{1-z_{ij}}{\eta\sigma_1^2}\bigg)\bigg]^{-1}
		\end{eqnarray*}
		
		\item
		$v_i|\bm{y},\bm{\beta},\bm{z},\sigma_1^2,\eta,\sigma_v^2,p_e\sim N\bigg(\varphi_i\displaystyle\sum_{j=1}^{n_i}(y_{ij}-x_{ij}^T\bm{\beta})\Big(\dfrac{z_{ij}}{\sigma_1^2}+\dfrac{1-z_{ij}}{\eta\sigma_1^2}\Big), \varphi_i\bigg)$
		\begin{eqnarray*}\text{where }\varphi_i=\bigg(\dfrac{1}{\sigma^2_v}+\displaystyle\sum_{j=1}^{n_i}\bigg(\dfrac{z_{ij}}{\sigma_1^2}+\dfrac{1-z_{ij}}{\eta\sigma^2_1}\bigg)\bigg)^{-1},i=1,...,m
		\end{eqnarray*}
		
		\item
		$z_{ij}|\bm{y},\bm{v},\bm{\beta},\sigma_1^2,\eta,\sigma_v^2,p_e \sim Bernoulli(p_{ij}^*), j=1,...,n, i=1,...,m$
		\begin{eqnarray*}
			\text{where }p^{*}_{ij}&=\dfrac{p_e\times\exp\bigg(-\dfrac{(y_{ij}-\bm{x_{ij}^T}\bm{\beta}-v_i)^2}{2\sigma_1^2}\bigg)}
			{p_e\times\exp\bigg(-\dfrac{(y_{ij}-\bm{x_{ij}^T}\bm{\beta}-v_i)^2}{2\sigma_1^2}\bigg)+\dfrac{1-p_e}{\sqrt{\eta}}\exp\bigg(-\dfrac{(y_{ij}-\bm{x_{ij}^T}\bm{\beta}-v_i)^2}{2\eta\sigma_1^2}\bigg)}
		\end{eqnarray*}
		
		\item
		$p_e|\bm{y},\bm{v},\bm{z},\bm{\beta},\sigma_1^2,\eta,\sigma_v^2\sim Beta\bigg(\displaystyle\sum_{i=1}^{m}\displaystyle\sum_{j=1}^{n_i}z_{ij}+1,
		\displaystyle\sum_{i=1}^{m}\displaystyle\sum_{j=1}^{n_i}(1-z_{ij})+1\bigg)\times I_{\{2^{-1}<p_e<1\}}$
		\bigskip\noindent In other words, we draw $p_e$ from a truncated Beta distribution with the first shape parameter $\displaystyle\sum_{i=1}^{m}\displaystyle\sum_{j=1}^{n_i}z_{ij}+1$, the second shape parameter $\displaystyle\sum_{i=1}^{m}\displaystyle\sum_{j=1}^{n_i}(1-z_{ij})+1$, and lower truncation point $2^{-1}$.
		
		\item
		$\dfrac{1}{\sigma_v^2}|\bm{y},\bm{v},\bm{z},\bm{\beta},\sigma_1^2,\eta,p_e\sim Gamma\bigg(\dfrac{m}{2}-1,\dfrac{1}{2}\displaystyle\sum_{i=1}^{m}v_i^2\bigg)$
		
		\item
		$\dfrac{1}{\sigma_1^2}|\bm{y},\bm{v},\bm{z},\bm{\beta},\sigma_v^2,\eta,p_e \sim \pi(\dfrac{1}{\sigma_1^2}|\bm{y},\bm{v},\bm{z},\bm{\beta},\sigma_v^2,\eta,p_e)\\[10pt]$
		\begin{eqnarray*}
			\propto Gamma\Bigg(\dfrac{n}{2},\dfrac{\displaystyle\sum_{i=1}^{m}\displaystyle\sum_{j=1}^{n_i}[(y_{ij}-\bm{x_{ij}^T}\bm{\beta}-v_i)^2(1+\eta z_{ij}-z_{ij})]}{2\eta}\Bigg)
		\end{eqnarray*}
		
		\item
		$\eta|\bm{y},\bm{v},\bm{z},\bm{\beta},\sigma_v^2,\sigma_1^2,p_e \sim \pi(\eta|\bm{y},\bm{v},\bm{z},\bm{\beta},\sigma_v^2,\sigma_1^2,p_e)$\\
		\begin{eqnarray*}
			&=\dfrac{\exp\bigg[ -\dfrac{1}{2}\displaystyle\sum_{i=1}^{m}\displaystyle\sum_{j=1}^{n_i}\bigg(\dfrac{(y_{ij}-\bm{x_{ij}^T}\bm{\beta}-v_i)^2}{\eta\sigma_1^2}(1-z_{ij})\bigg)\bigg]}{(\eta)^{\big(\dfrac{1}{2}\displaystyle\sum_{i=1}^{m}\displaystyle\sum_{j=1}^{n_i}(1-z_{ij})+1\big)}} \times \dfrac{\eta}{(1+\eta)^2}
		\end{eqnarray*}
	\end{enumerate}
	
	
\end{document}